\documentclass[12pt,  sagev,times]{sagej}
\usepackage{geometry} 
\usepackage{moreverb,url}
\usepackage[colorlinks,bookmarksopen,bookmarksnumbered,citecolor=red,urlcolor=red]{hyperref}
\usepackage{xcolor}

\newcommand\BibTeX{{\rmfamily B\kern-.05em \textsc{i\kern-.025em b}\kern-.08em
T\kern-.1667em\lower.7ex\hbox{E}\kern-.125emX}}

\usepackage{multirow}
\usepackage{soul}

\begin{document}

\runninghead{Menzies, Saint-Hilary and Mozgunov}

\title{A Comparison of Various Aggregation Functions in Multi-Criteria Decision Analysis for Drug Benefit-Risk Assessment}

\author{Tom Menzies\affilnum{1,2}, Gaelle Saint-Hilary\affilnum{3,4} and Pavel Mozgunov\affilnum{5}}

\affiliation{\affilnum{1}Clinical Trials Research Unit, Leeds Institute of Clinical Trials Research, University of Leeds, Leeds, UK \\
\affilnum{2}Department of Mathematics and Statistics, Lancaster University, Lancaster, UK  \\
\affilnum{3}Department of Biostatistics, Institut de Recherches Internationales Servier (IRIS), Suresnes, France\\
\affilnum{4}Dipartimento di Scienze Matematiche (DISMA) Giuseppe Luigi Lagrange, Politecnico di Torino, Torino, Italy\\
\affilnum{5}Medical and Pharmaceutical Statistics Research Unit, Department of Mathematics and Statistics, Lancaster University, Lancaster, UK }

\corrauth{Tom Menzies, Clinical Trials Research Unit,
University of Leeds,
Leeds,
LS2 9JT, UK.}
\email{t.menzies@leeds.ac.uk}

\begin{abstract}
Multi-criteria decision analysis (MCDA) is a quantitative approach to the drug benefit-risk assessment (BRA) which allows for consistent comparisons by summarising all benefits and risks in a single score. The MCDA consists of several components, one of which is the utility (or loss) score function that defines how benefits and risks are aggregated into a single quantity. While a linear utility score is one of the most widely used approach in BRA, it is recognised that it can result in counter-intuitive decisions, for example, recommending a treatment with extremely low benefits or high risks. To overcome this problem, alternative approaches to the scores construction, namely, product, multi-linear and Scale Loss Score models, were suggested. However, to date, the majority of arguments concerning the differences implied by these models are heuristic. In this work, we consider four models to calculate the aggregated utility/loss scores and compared their performance in an extensive simulation study over many different scenarios, and in a case study. It is found that the product and Scale Loss Score models provide more intuitive treatment recommendation decisions in the majority of scenarios compared to the linear and multi-linear models, and are more robust to the correlation in the criteria.
\end{abstract}


\maketitle

\section{Introduction}
\label{Introduction}

The benefit-risk analysis of a treatment consists of balancing its favourable therapeutic effects versus adverse reactions it may induce~\citep{R1}. This is a process which drug regulatory authorities, such as EMA~\cite{R18} and FDA~\cite{R19} use when deciding whether a treatment should be recommended. Benefit-risk assessment (BRA) is mostly performed in a qualitative way~\cite{R21}. However, this approach has been criticised for a lack of transparency behind the final outcome, in part due to large amounts of data considered for this assessment, and the differing opinions on what this data means. To counter this, quantitative approaches ensuring continuity and consistency across drug BRA, and making the decisions easier to justify and to communicate, were proposed~\cite{R22,R23}.

While there is a number of methods to conduct the quantitative BRA, the multi-criteria decision analysis (MCDA) has been particularly recommended by many expert groups in the field~\citep{R3}\citep{R20}\citep{R31}\citep{R32}. MCDA provides a single score (a utility or loss score) for a treatment, which summarises all the benefits and risks induced by the treatment in question. These scores are then used to compare the treatments and to guide the recommendation of therapies over others. 

Mussen \textit{et al.}~\cite{R3} proposed to use a linear aggregation model in the MCDA, which takes into account all main benefits and risks associated with a treatment (as well as their relative importance) to generate a treatment utility score by taking a linear combination of all criteria. This utility score is then compared against the utility score of a competing treatment, and that with the highest score is recommended. This model appealed for numerous reasons, one of which was its simplicity. The proposed method, however, was deterministic, point estimates of the benefit and risk criteria were used, and no uncertainty around these estimates was considered. Yet, uncertainty and variance are expected in treatments' performances, and must therefore be accounted for in the decision-making.

To resolve this shortcoming, probabilistic MCDA (pMCDA)~\citep{R9} that accounts for the variability of the criteria through a Bayesian approach was proposed. Generalisations of pMCDA for the case of uncertainty in the relative importance of the criteria were developed, named stochastic multi-criteria acceptability analysis (SMAA)~\cite{R14} or Dirichlet SMAA~\cite{R4}. However, it was acknowledged that by accounting for several sources of uncertainty, these models become more complex and should be used primarily for the sensitivity analysis.

All the works discussed above concern a linear model for aggregation of the criteria, which is thought to be primarily due to its wider application in practice rather than its properties. One argument against the linear model is that a treatment which has either no benefit or extreme risk could be recommended over other alternatives without such extreme characteristics.~\citep{R6} \citep{R33} \citep{R34}. In addition, the linearity implies that the relative tolerance in the toxicity increase is
constant for all levels of benefit that might not be the case for a number of clinical settings. To address these points, a Scale Loss Score (SLoS) model was developed. This model made it impossible for treatments with no benefit or extremely high risk be recommended. It also incorporates a decreasing level of risk tolerance relative to the benefits: where an increase in risk is more tolerated when benefit improves from ``very low" to ``moderate" compared to an increase from ``moderate" to ``very high". SLoS model resulted in similar recommendations to the linear MCDA model when the one treatment is strictly preferred to another (i.e. has both lower risk and higher benefit), but resulted in more intuitive recommendations if one of the treatments has either extremely low benefit or extremely high risk. 

Whilst other methods are discussed in the literature, the only application of a non-linear BRA model to the medical field is made by Saint-Hilary et al \citep{R6}, and this only compares the linear and SLoS models. This paper shall build on this comparison by introducing various different aggregation models (AM) to analyse how each work compared to the other in the medical field (by conduction a case study and a simulation study), and allow an informed decision to be made as to which one should be used using the results of an extensive and comprehensive simulations study over a number of clinical scenarios. We will also use a case study to demonstrate the implication of the choice of AM on the actual decision-making using the MCDA.

The rest of the paper proceeds as follows. The general MCDA methodology, the four different aggregation models considered, linear, product, multi-linear and SLoS, and the choice of the weights for them are given in Section~2. In Section~3, we revisit a case study conducted conducted by Nemeroff~\citep{R15} looking at the effects of Venlafaxine, Fluoxetine and a placebo on depression, applying the various aggregation models to a given dataset. In Section~4, a comprehensive simulation study comparing the four aggregation models in many different scenarios is presented, as well as the effects any correlation between criteria may have. We conclude with a discussion in Section 5.

\section{Methodology}
\label{Methodology}

All of the aggregation models (referred to as to ``models'' below) considered in this work are all classified within the MCDA family - they aggregate the information about benefits and risks in a single (utility or loss) score. Therefore, we would refer to each of the approaches by their models for the computation of the score. Below, we outline the general MCDA framework for the construction of a score using an arbitrary model.
We consider the MCDA taking into account the variability of estimates, pMCDA~\cite{R9}.
\subsection{Setting}

Consider $m$ treatments (indexed by $i$) which are assessed on $n$ criteria (indexed by $j$). To ensure continuity, we use the same notations as those of Saint-Hilary \textit{et al.}~\cite{R6}: 
\begin{itemize}
\item $\xi_{i,j}$ is the performance of treatment $i$ on criterion $j$, so that treatment $i$ is characterised by a vector showing how it performed on each criterion: $\boldsymbol{\xi_{i,j}}$ = ($\xi_{i,1},......,\xi_{in}$).

\item The monotonically increasing partial value functions $0\leq u_j(\cdot) \leq 1$ are used to normalise the criterion performances. Let $\xi'_{j}$ and $\xi''_{j}$ be the most and the least preferable values, then $u_j(\xi''_{j})=0$ and $u_j(\xi'_{j})=1$. The inequality $u_j(\xi_{ij})>u_j(\xi_{hj})$ indicates that the performance of the treatment $i$ is preferred to the performance of the treatment $h$ on criterion $j$. In this work, we focus
 on linear partial value functions, one of the most common choice in treatment benefit-risk assessment \cite{R22, R3, R14, R27, R9} that can be written as
\begin{equation}
u_j(\xi_{ij})=\frac{\xi_{ij}-\xi''_{j}}{\xi'_{j}-\xi''_{j}}.
\label{eq:partial}
\end{equation}

\item The weights indicating the relative importance of the criteria are known constants denoted by $w_j$. The vector of weights used for the analysis is denoted by $\boldsymbol{w}=\left(w_1, ..., w_n\right)$.

\item The MCDA utility or loss scores of treatment $i$ are obtained as
$$u(\boldsymbol{\xi}_i,\boldsymbol{w}):=u\left(w_j, u_j(\xi_{ij})\right), ~~j=1,.....,n
$$ and
$$
l(\boldsymbol{\xi}_i,\boldsymbol{w}):=l\left(w_j, u_j(\xi_{ij})\right), ~~j=1,.....,n
$$ respectively, where $u\left(\cdot\right)$ and $l\left(\cdot\right)$ are the functions specifying how the criteria should be summarised in a single score, and are referred to as ``aggregation models''. The impact of this model's choice on the performance of treatment recommendation is the focus on this work.
The higher the utility score, or lower the loss score, the more preferable the benefit-risk ratio. Then, the comparison of treatments $i$ and $h$ is based on

\begin{equation}
\Delta u(\boldsymbol{\xi}_i,\boldsymbol{\xi}_{h},\boldsymbol{w}) := u(\boldsymbol{\xi}_i,\boldsymbol{w}) - u(\boldsymbol{\xi}_{h},\boldsymbol{w})   
\label{eq:difference}
\end{equation}
or 
\begin{equation}
\Delta l(\boldsymbol{\xi}_i,\boldsymbol{\xi}_{h},\boldsymbol{w}) := l(\boldsymbol{\xi}_i,\boldsymbol{w}) - l(\boldsymbol{\xi}_{h},\boldsymbol{w}).
\label{eq:difference-loss}
\end{equation}
\end{itemize}
Within a Bayesian approach, the utility score $u(\boldsymbol{\xi}_i,\boldsymbol{w})$ and the loss score $l(\boldsymbol{\xi}_i,\boldsymbol{w})$ are random variables having a prior distribution. Given observed outcomes $\mathbf{x_{i}}=(x_{i1},\ldots,x_{in})$ and $\mathbf{x_{h}}=(x_{h1},\ldots,x_{hn})$ (corresponding to treatment performances $\boldsymbol{\xi}_{i}$ and $\boldsymbol{\xi}_{h}$, respectively) for $i$ and $h$, one can obtain the posterior distribution of $\Delta u(\boldsymbol{\xi}_i,\boldsymbol{\xi}_{h},\boldsymbol{w})$ or $\Delta l(\boldsymbol{\xi}_i,\boldsymbol{\xi}_{h},\boldsymbol{w})$, respectively. The inference is based on the complete posterior distribution and the conclusion on the benefit-risk balance is supported by the probability of treatment $i$ to have a greater utility score (or smaller loss score) than treatment~$h$:
\begin{equation}\label{eq:probability}
\mathcal{P}_u^{ih}=\mathbb{P}(\Delta u(\boldsymbol{\xi}_i,\boldsymbol{\xi}_{h},\boldsymbol{w})>0 \mid \mathbf{x_{i}},\mathbf{x_{h}}).
\end{equation}
or
\begin{equation}\label{eq:probability-loss}
\mathcal{P}_l^{ih}=\mathbb{P}(\Delta l(\boldsymbol{\xi}_i,\boldsymbol{\xi}_{h},\boldsymbol{w})<0 \mid \mathbf{x_{i}},\mathbf{x_{h}}).
\end{equation}

The probabilities~\eqref{eq:probability} or ~\eqref{eq:probability-loss} are used to guide a decision on taking/dropping a treatment. A possible way to formalise the decision based on this probability is to compare it to a threshold confidence level $ 0.5 \leq \psi \leq 1$. Then, $\mathcal{P}_u^{ih} > \psi$ (or $\mathcal{P}_l^{ih} > \psi$) would mean that one has enough evidence to say that treatment $i$ has a better benefit-risk balance than $h$ with a level of confidence $\psi$. Note that $\mathcal{P}_u^{ih}=0.5$ (and $\mathcal{P}_l^{ih}=0.5$) corresponds to the case where the benefit-risk profiles of $i$ and $h$ are equal according to the corresponding MCDA model.

\subsection{Aggregation Models}
\label{Aggregation Models}
Below, we consider four specific forms of aggregation models, namely, linear, product, multi-linear, and Scale Loss Score, that were argued by various authors to be used in the MCDA to support decision-making 

\subsubsection{Linear Model}
\label{Linear Model}

A linear aggregation of treatment's effects on benefits and risks remains the most common choice for the treatment development \cite{R3, R14, R28, R27, R4}. Under the linear model, the utility score is computed as 
\begin{equation}
u^{{{L}}}(\boldsymbol{\xi_{i}},\boldsymbol{w}^{{L}}) := \sum_{j=1}^{n}w_{j}^{{L}} u_{j}(\xi_{i,j})
\label{eq:linear}
\end{equation}
where $w_{j}^{L} > 0$ $\ \forall j$ and $\sum_{j=1}^n w_{j}^{L}=1$, the superscript $L$ referring to the linear model. The expression~\eqref{eq:linear} is used in Equation~\eqref{eq:difference} and Equation~\eqref{eq:probability} to compare the associated linear scores for a pair of treatments. 

As an illustration of all considered aggregation models, we will use the following example with two criteria: one benefit indexed by $1$, one risk indexed by $2$. The linear utility score for treatment $i$ at fixed parameter values $\theta_{i1}$, $\theta_{i2}$ takes the form
\begin{equation}
u^{L}(\theta_{i1},\theta_{i2},w^{L}):= w^{L}u_1(\theta_{i1})+(1-w^{L})u_2(\theta_{i2}).
\label{eq:exampleL}
\end{equation}

\begin{figure}[h]
\begin{center}
\includegraphics[width=17.6cm]{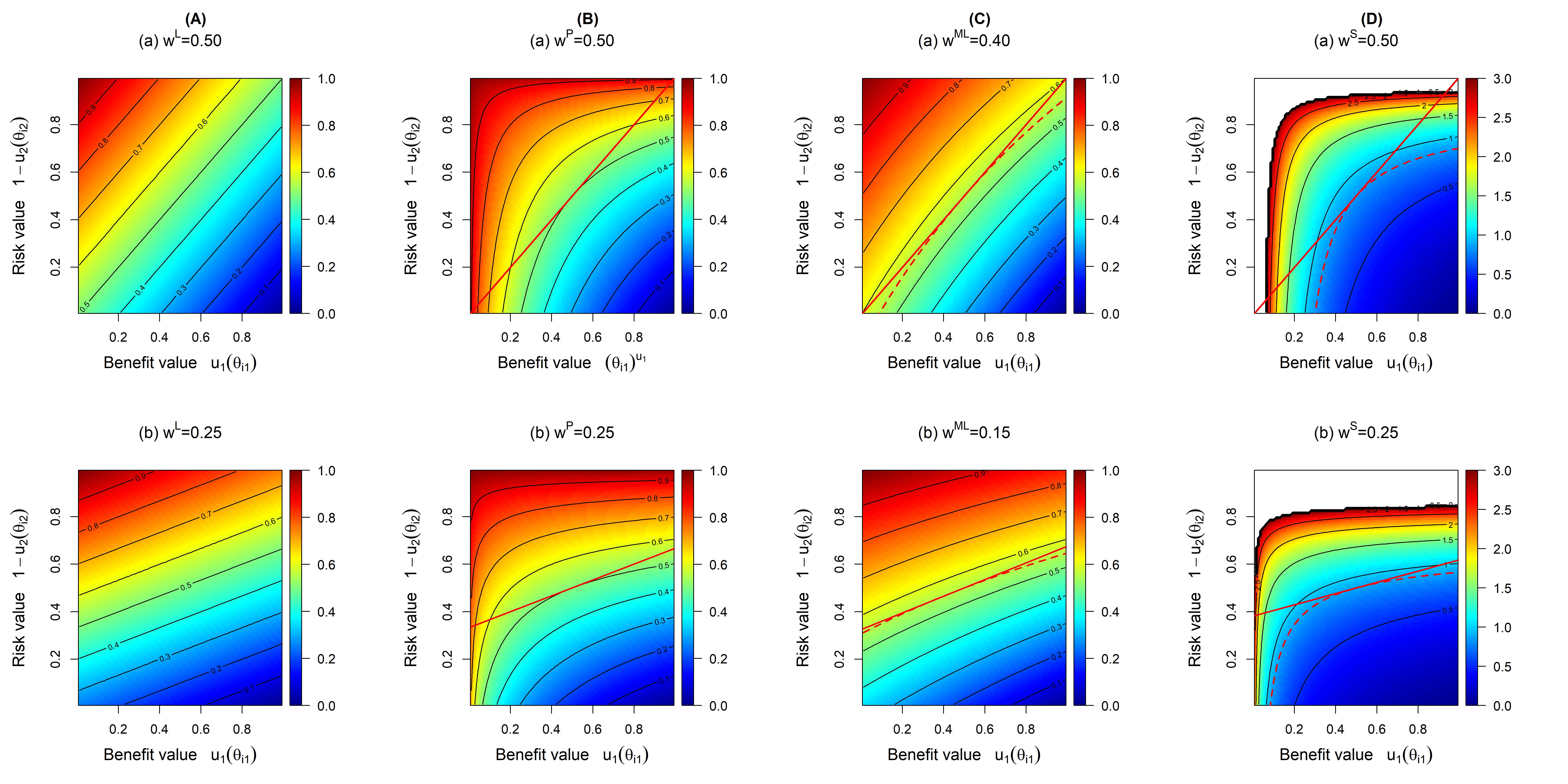}
\end{center}
\vspace{-2em}
\caption{Contour plots for Linear (A), Product (B), Multi-Linear (C), and SLoS (D) models with (i) two equally important criteria (top row), and (ii) the risk criterion being twice as important (on average for non-linear model) as the benefit criterion (bottom row). Red lines on Panels B--D represents the tangents at the middle point (0.5,0.5).}
\label{fig:contour1}
\end{figure}

As values $u_1(\theta_{i1}), u_2(\theta_{i2}) \in (0,1) $, one can interpret $u_1(\theta_{i1})$ as a probability of benefit and $1-u_2(\theta_{i2})$ as a probability of risk. This utility score can be transformed into a loss score by subtracting it from one:
\begin{equation}
l^{L}(\theta_{i1},\theta_{i2},w^{L}) := 1-u^{L}(\theta_{i1},\theta_{i2},w^{L})
\label{eq:LinLossL}
\end{equation}
We do this as, historically, the concept of a loss function is preferred both in statistical decision theory and Bayesian analysis for parameter estimation.\citep{R7} The contours of equal linear loss score for all values of $u_1(\theta_{i1})$ and $(1-u_2(\theta_{i2}))$ are given in Panel (A) of Figure \ref{fig:contour1} using $w^L=0.5$ (top row) and $w^L=0.25$ (bottom row).

The contours represent the loss score for each benefit-risk pair. Lower values of $l^{L}(\theta_{i1},\theta_{i2},w^{L})$ correspond to better treatment benefit-risk profiles. It is minimised (right bottom corner) when the maximum possible benefit is reached ($u_{1}(\theta_{i1}$) = 1) with no risk (1-$u_{2}(\theta_{i2}$) = 0). The contours are linear, with a constant slope $w^{L}$/(1-$w^{L}$). This implies that if one treatment has an increased probability of risk of x$\%$ compared to another, its benefit probability should be increased by (1-$w^{L}$)/$w^{L}$ $\times$ x$\%$ to have the same utility score, and this holds for all values of benefit and risk. This figure allows for an illustration of the penalisation of various benefit-risk criteria and for an illustrative comparison between treatments with different criteria. For example, any pairwise comparison that lies on a contour line shows that the two treatments are seen as equal.

The major advantage of the linear model is its intuitive interpretation: a poor efficacy can be compensated by a good safety, and vice-versa. However, the linear utility score can result in the recommendation of highly unsafe or poorly effective treatment \cite{R29, R23} and, consequently, in a counter-intuitive conclusion. Moreover, the linearity implies that the relative tolerance in the toxicity increase is constant for all levels of benefit~\cite{R6}. These pitfalls could be avoided (or at least reduced) by using non-linear models \cite{R23, R11}. Specifically, Saint-Hilary \textit{et  al.}~\cite{R6} advocated introducing two principles a desirable benefit-risk analysis aggregation model should have:

\begin{enumerate}
\item One is not interested in treatments with extremely low levels of benefit or extremely high levels of risks (regardless of how the treatment performs on other criteria);
\item For an equivalent absolute increase in benefit, one can tolerate a larger risk increase if the amount of benefit is small than if it is high.
\end{enumerate}
Below, we consider three models having one or both of these properties.

\subsubsection{Product Model}
\label{Product Model}

A multiplicative aggregation (known as a product model) is an alternative method of comparing treatment's effects on benefits and risks\citep{R8}. Under the product model, the utility score is computed as 
\begin{equation}
u^{{{P}}}(\boldsymbol{\xi_{i}},\boldsymbol{w}^{{P}}) := \prod_{j=1}^{n} u_{j}(\xi_{i,j})^{w_{j}^{{P}}}
\label{eq:product}
\end{equation}
where the superscript $P$ refers to the product model. The expression~\eqref{eq:product} is used in Equation~\eqref{eq:difference} and Equation~\eqref{eq:probability} to compare the associated product scores for a pair of treatments. 

The product utility score for treatment $i$ with two criteria at fixed parameter values $\theta_{i1}$, $\theta_{i2}$ takes the form
\begin{equation}
u^{P}(\theta_{i1},\theta_{i2},w^{P}):= u_1(\theta_{i1})^{w^{P}} \times u_2(\theta_{i2})^{(1-w^{P})}.
\label{eq:exampleP}
\end{equation}

Similarly as for the linear model, this utility score can be transformed into a loss score by subtracting it from one:
\begin{equation}
l^{P}(\theta_{i1},\theta_{i2},w^{P}) := 1-u^{P}(\theta_{i1},\theta_{i2},w^{P})
\label{eq:LinLossP}
\end{equation}
The contours of equal product loss score for all values of $u_1(\theta_{i1})$ and $(1-u_2(\theta_{i2}))$ are given in Panel (B) of Figure \ref{fig:contour1} using $w^P=0.5$ (top row) and $w^P=0.25$ (bottom row). 

One advantage the product model has over the linear model is that it cannot recommend treatments with either zero benefit or extreme risk. This is because either of these two options would result in a score of zero for the utility function, and as such would make it impossible for such a treatment to be recommended.
The contour lines in Panel (B) in Figure \ref{fig:contour1} demonstrate how the product model penalises undesirable values compared to the linear model. These contours are curved, and are bunched together tightest at points where benefit values are low and where risk values are high. This shows how the penalisation differs this model from the linear model, as under the linear model, an increase/decrease in benefit-risk is treated equally regardless of the marginal values of these criteria, whereas the values of these criteria often have an effect on our decision making under the product model.

\subsubsection{Multi-Linear Model}
\label{Multi-Linear Model}

A multi-linear model for the aggregation of treatments' benefits and risks provides a one more alternative for the comparison of two treatments~\cite{R11}. This model can be seen as attempt to combine the linear and product model. Under the multi-linear model, the utility score is computed as 
\begin{equation}
\begin{array}{c}
u^{{{ML}}}(\boldsymbol{\xi_{i}},\boldsymbol{w}^{{ML}}) := \sum_{j=1}^{n}w_{j}^{{ML}} u_{j}(\xi_{i,j}) + 
\sum_{j=1, k>j}^{n} w^{ML}_{j,k} u_{j}(\xi_{i,j})u_{k}(\xi_{i,k}) + \\ 
\sum_{j=1, l>k>j}^{n} w^{ML}_{j,k,l} u_{j}(\xi_{i,j})u_{k}(\xi_{i,k})u_{l}(\xi_{i,l}) + ...  
... + w^{ML}_{1,2,....,n} u_{1}(\xi_{i,1})u_{2}(\xi_{i,2})....u_{n}(\xi_{i,n})
\end{array}
\label{eq:multilinear}
\end{equation}
where the superscript $ML$ refers to the multi-linear model, and the weight criteria $w^{ML}_{i,j,...}$ refer to the weight criteria given to the interaction term between criteria $i,j,...$. We require all the weights in the ML model to sum up to 1. The expression~\eqref{eq:multilinear} is used in Equation~\eqref{eq:difference} and Equation~\eqref{eq:probability} to compare the associated multi-linear scores for a pair of treatments.

Considering the example with two criteria, the multi-linear utility score for treatment $i$ at fixed parameter values $\theta_{i1}$, $\theta_{i2}$ takes the form 
\begin{equation}
u^{ML}(\theta_{i1},\theta_{i2},w_1^{{ML}},w_2^{{ML}},w_{1,2}^{{ML}}):= w_1^{{ML}}u_1(\theta_{i1})+w_2^{{ML}}u_2(\theta_{i2})+w^{ML}_{1,2}( u_1(\theta_{i1})u_1(\theta_{i2})).
\label{eq:exampleML}
\end{equation}
Note that the even under the constraint of the sum of the weights to be equal to one, there is one more weight parameter than for the linear and product models. This immediately can make the weight elicitation procedure more involving for all stakeholders. To link the weights of the ML model with the rest of the competing approaches (see more details in Section 2.3), we set up one more constraint, so that the number of weight parameters is the same in all considered model (for the purpose of the comparison in this manuscript). Specifically, we fix $w^{ML}_{1,2}=c$ where $0\leq c \leq 1$, implying that we fix the effect of the interaction term. 
Similarly as for the linear and product models, this utility score can be transformed into a loss score by subtracting it from one:
\begin{equation}
l^{ML}(\theta_{i1},\theta_{i2},w^{ML}) := 1-u^{ML}(\theta_{i1},\theta_{i2},w^{ML})
\label{eq:LinLossML}
\end{equation}
The contours of equal linear loss score for all values of $u_1(\theta_{i1})$ and $(1-u_2(\theta_{i2}))$, $c=0.20$ are given in Panel (C) of Figure \ref{fig:contour1} using $w_1^{ML}=0.40$ (top row) and $w_1^{ML}=0.15$ (bottom row).

The contour lines demonstrate the almost linear trade-off between benefit and risk, but that there is a slight curvature (which becomes more prominent as it moves further away from more desirable values), indicating a moderate penalisation of extreme values. This shows that while this model attempts to penalise the undesirable criteria values, this effect does not seem to be as strong as in the product model, admittedly due to the chosen value of the weight, $w^{ML}_{1,2}$, given to the interaction term . A moderate level of penalisation for the chosen value of the weight corresponding to the interaction term allows for treatments to be recommended when there is no benefit or extreme risk, as is the case in the linear model. The more the weight of the interaction terms, the less likely this would happen.

\subsubsection{Scale Loss Score (SLoS) Model}
\label{Scale Loss Score (SLoS) Model}
An alternative to the models proposed above is the Scale Loss Score (SLoS) model, which was proposed by Saint-Hilary \textit{et al.}~\cite{R6} to satisfy the two desirable properties for an aggregation method. First of all, in contrast to the three models above, SLoS considers a loss score, rather than a utility score, as the output. Therefore, lower values are more desirable. Under the SLoS model, the loss score is computed as 
\begin{equation}
l^{{{S}}}(\boldsymbol{\xi_{i}},\boldsymbol{w}^{{S}}) := \sum_{j=1}^{n} \bigg(\frac{1}{ u_{j}(\xi_{i,j})}\bigg)^{w_{j}^{{S}}}
\label{eq:slos}
\end{equation}
where the superscript $S$ refers to the SLoS model. The expression~\eqref{eq:slos} is used in Equation~\eqref{eq:difference-loss} and Equation~\eqref{eq:probability-loss} to compare the associated SLoS scores for a pair of treatments. 

Coming back to the example with two criteria, the loss score for treatment $i$ at fixed parameter values $\theta_{i1}$, $\theta_{i2}$ takes the form
\begin{equation}
l^{s}(\theta_{i1},\theta_{i2},w^{{S}}):= \bigg(\frac{1}{u_1(\theta_{i1})}\bigg)^{w^{{S}}}+\bigg(\frac{1}{u_2(\theta_{i2})}\bigg)^{(1-w^{{S}})}.
\label{eq:exampleS}
\end{equation}

The contours of equal scale loss score for all values of $u_1(\theta_{i1})$ and $(1-u_2(\theta_{i2}))$ are given in Panel (D) of Figure \ref{fig:contour1} using $w^S=0.5$ (top row) and $w^S=0.25$ (bottom row). 

As is the case with the product model, this penalisation makes it impossible for treatments with either no benefit or extreme risk to be recommended over other potential treatments, compared to the linear and multi-linear models (which can recommend such treatments). This is because a treatment that had either of these would return a loss score of infinity (regardless of the values of any other criteria) and would therefore be non-recommendable. On the Figure, the white colour at extreme undesirable values (either very low benefit or very high risk) corresponds to very high to infinite loss scores and demonstrate the penalisation effect.

Even when the contour plots in Figure \ref{fig:contour1} concern the same \textit{values} of weights in the models, the weights themselves are different in each model (represented by different indices). Therefore, when to provide a fair comparison of these models, it is important to ensure that the models carry (approximately) the same relative importance of the criteria defined through the slope of the contour lines. We propose an approach to match the relative importance of the models below.

\subsection{Weight Elicitation and Mapping}
\label{Weight Elicitation and Mapping} 

Methods for quantifying subjective preferences, for example, Discrete Choice Experiment and Swing-Weighting, have been widely studied in the literature~\cite{R23,R3, R25, R26}. Applied to drug BRA, the majority of the weight elicitation methods concern the linear model. In the linear model framework, the weight assigned to one criterion is interpreted as a scaling factor which relates one increment on this criterion to increments on all other criteria.

Note that each of the aggregation models use the individual weights, $w^{L}, w^{P}, w^{ML}$, and $w^{S}$. However, in the actual analysis, regardless of the aggregation model used, one can expect only one underlying level of the relative importance of the considered benefit and risk criteria, as the stakeholders' preferences between the criteria should not depend on the methodology used for the decision-making. Therefore, it is crucial to make sure when applying different models to the same problem that they reflect the same stakeholders' preferences. We adapt the approach proposed by Saint-Hilary \textit{et al.}~\cite{R6} to achieve that. Since comprehensive work has been published and is currently being continued on the weight elicitation for the linear model, we will map the weights $w^{L}_j$ (hypothetically) elicited for the linear model to the weights $w^{P}, w^{ML}$, and $w^{S}$ such that they reflect the same trade-off preferences between the criteria.

\subsubsection{Mapping for Two Criteria}
\label{Mapping for Two Criteria}

As described in Saint-Hilary \textit{et al.}~\cite{R6}, formally, the trade-off between the criteria could be represented by the slope of the tangent of the contour lines where the contour line passes through the point (0.5, 0.5) (see the red lines in the contour plot of Panels B-D in Figure \ref{fig:contour1}). Therefore, the expressions for the mapping of the linear weight to the competitive models are found through the equality of the slopes of the tangents to the corresponding contour lines.


We start from the setting with two criteria. As stated above, even for the two criteria setting, the multi-linear model requires one more weight to be specified. Therefore, we impose a constraint on the weight corresponding to the interaction term to obtain the unique solution for the mapped weight $w^{ML}$, specifically $w^{ML}_{1,2}=1-w_1^{ML}-w_2^{ML}=c$, where $0 \leq c \leq 1$. Note that for $c=0$, the multi-linear model reduces to the linear one, and for $c=1$ it becomes the product of the two criteria values. 

Using the utility/loss scores $z^{{P}},z^{{ML}},z^{{S}}$ obtained at point $(u_1(\theta_{i1}),u_2(\theta_{i2}))$, the expressions of the equality of the tangents with two criteria take the form

\begin{equation}
\begin{array}{ccc}
\frac{w^{{L}}}{1-w^{{L}}} & = & \frac{w^{{P}}}{1-w^{{P}}} \bigg(\frac{1}{u_{1}(\theta_{i,1})}\bigg)\bigg(\frac{z^{{P}}}{u_{1}(\theta_{i,1})^{w^{_{P}}}}\bigg)^{\frac{1}{1-w^{{P}}}} ,\\
\frac{w^{{L}}}{1-w^{{L}}} & = & \frac{w_{1}^{ML}w_{2}^{ML}+z^{ML}-z^{ML}(1-c)}{\big{(}w_{2}^{ML}+cu_{1}(\theta_{i,1})\big{)}^2}, \\
\frac{w^{{L}}}{1-w^{{L}}} & = & \frac{{w^S}}{1-{w^S}}\left(z^S-u_1(\theta_{i1})^{-{w^S}}\right)^{\frac{{w^S}-2}{1-{w^S}} } \ \times \ u_1(\theta_{i1})^{-({w^S}+1)}.
\end{array}
\end{equation}
where the slope for the linear model is given in the left hand size, and the slopes for the product, multi-linear and SLoS models are given in the right hand side, respectively. 

Note, however, that the slope of the tangent of the contours for the linear model are \textit{constant} for all values of parameters and defined by the weights $w^{L}_j$ only, while the slopes for the competitive models change with the values of the criteria. For the purpose for the weights mapping, we would interpret $w^{L}_j$ as an \textit{average} relative importance of each criterion over the others, and would match the slopes of the tangents to the corresponding contours in the middle point, $u_1(\theta_{i1})=u_2(\theta_{i2})=0.5$~\cite{R6}. Then, the equalities above reduce to
\begin{equation}
\begin{array}{ccc}
w^{{L}} & = & w^{{P}} ,\\
{w^{{L}}} & = & {w^{{ML}}} + c/2, \\
\frac{w^{{L}}}{1-w^{{L}}} & = & \frac{w^S}{1-w^S}\ .\ 2^{(2w^S-1)}.
\end{array}
\label{eq:mapping}
\end{equation}

\begin{figure}[h]
\begin{center}
\includegraphics[width=10cm]{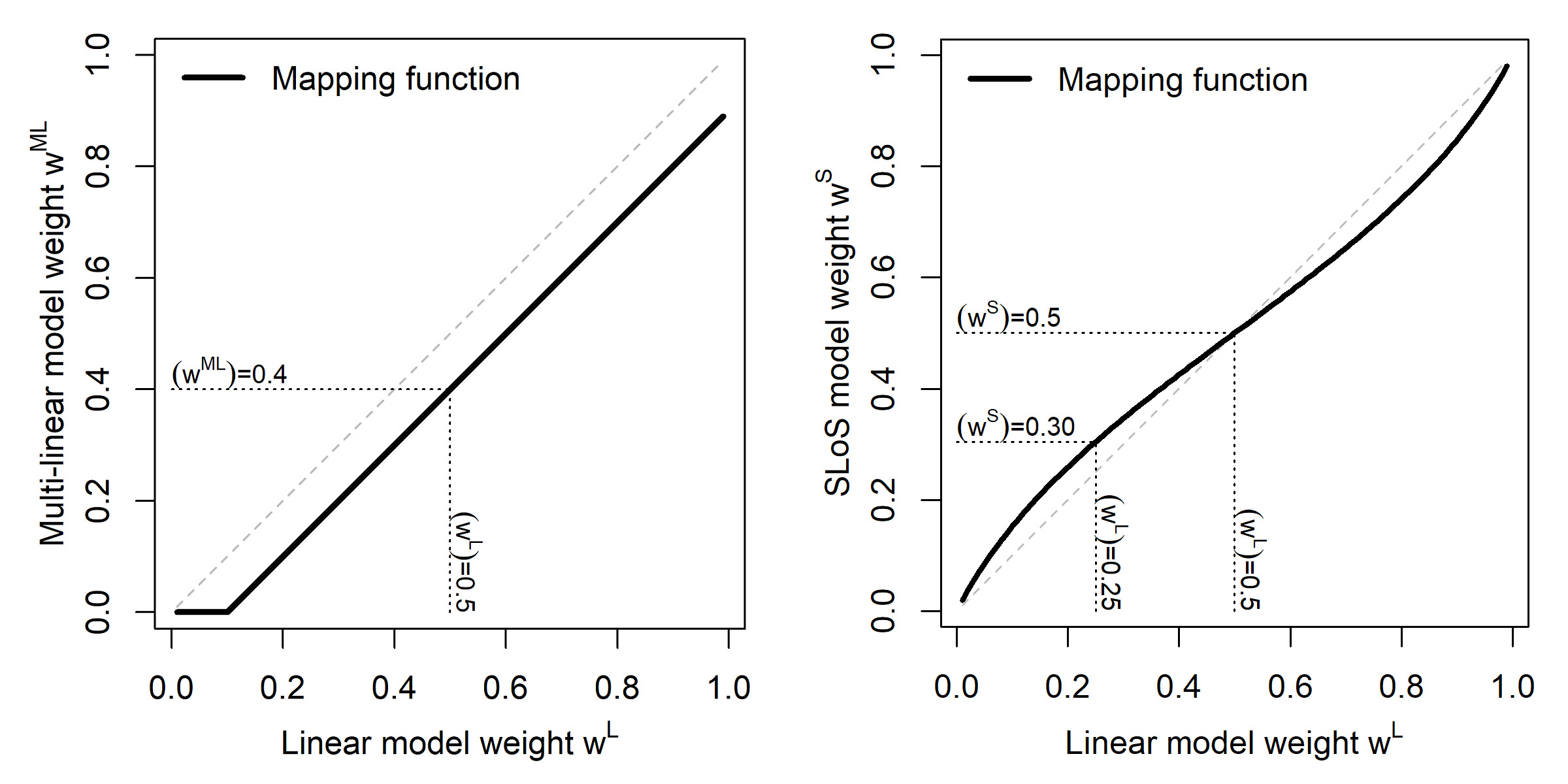} 
\end{center}
\vspace{-2em}
\caption{Weight mapping from the linear model to the multi-linear model (left) and to the SLoS model (right).}
\label{fig:mapping}
\end{figure}

Therefore, the product weight coincides with the linear weight in the given middle mapping point. For the SLoS model, the weight mapping does not have an analytical solution, but the approximate value of ${w}^{S}$ can be obtained by line search. Figure \ref{fig:mapping} shows the mapping from the linear model to the multi-linear and SLoS models. It demonstrates how the value for the linear model (x-axis) can be used to find the respective weights for the multi-linear and SLoS models on the y-axes.

One can note that for the multi-linear model, the proposed mapping process may result in the obtained negative mapped values of weight. This is because of how the weight mapping function is elicited in the two criteria case: if the value of a weight under the linear model is less than half the value of $c$, then this will map to a negative value (which, in theory, gives our criteria a negative importance - which is impossible) to reflect the same relative importance as induced by the linear model. Intuitively, if the interaction terms already contributes more to the importance of the one of the criterion in the interaction, the model needs to subtract the ``excessive'' importance from the weight corresponding to this criterion standing alone. Whilst this effect can be negated by setting an upper limit of the values $c$ can take, this in term limits the effect the interaction terms have, and can make the model more similar to the linear model. This is demonstrated in Figure \ref{fig:mapping} for $c=0.2$, where any weights for the linear model that are given a value of 0.1 or less would be mapped to 0 in the multi-linear model, rather than a negative value.

Proof for the above workings is given in the Supplementary Material.

\subsubsection{Mapping for Setting with More Than Two Criteria}
\label{Mapping for Setting with More Than Two Criteria}
The derivation above concerns the setting with two criteria only but could be directly extended for the product and SLoS models. Specifically, one can apply the proposed mapping function to each of the weights in the setting with more than two criteria marginally. This would imply that the weights are mapped with respect to the importance of all other criteria rather than a single benefit (or risk)~\cite{R6}.

The extension for the multi-linear model, however, is less straightforward. Generally, it would be a much more involving procedure to elicit weights for all the interactions terms as their number increases noticeably if more than two criteria are considered. Specifically, in the case study considered in Section~3, there are 4 criteria resulting in 11 interaction terms. Following the two criteria setting, we suggest to fix the total weight attributed to \textit{all} the interactions to be equal to $c=0.2$. Then, the ML model for the setting with 4 criteria takes the form
\begin{equation}
\begin{array}{c}
u^{{{ML}}}(\boldsymbol{\xi_{i}},\boldsymbol{w}^{{ML}}) := \sum_{j=1}^{n}w_{j}^{{ML}} u_{j}(\xi_{i,j})  + \frac{c}{2^n-n-1}\sum_{j=1, k>j}^{n} u_{j}(\xi_{i,j})u_{k}(\xi_{i,k}) + \\ \frac{c}{2^n-n-1}\sum_{j=1, l>k>j}^{n} u_{j}(\xi_{i,j})u_{k}(\xi_{i,k})u_{l}(\xi_{i,l}) + ... 
... + \frac{c}{2^n-n-1}u_{1}(\xi_{i,1})u_{2}(\xi_{i,2})....u_{n}(\xi_{i,n})
\end{array}
\label{eq:multilinear3}
\end{equation}
where the fraction $\frac{c}{2^n-n-1}$ ensures that the sum of all the interaction terms equals $c$ and this is split equally between all interaction terms. To calculate the individual weights $w_{j}, j=1,\ldots,n$, again, a mapping to the linear weights can be used. In order for the weights to sum up to 1, the transformation ${w^{{ML}}} =  {w^{{ML}}} - c/n$ could be applied. For $n=2$, this translates into the corresponding mapping in Equation~\ref{eq:mapping}. While this procedure does not guarantee the equality of the slopes of the tangents, it, however, emphasises the potential challenge associated with the use of the multi-linear model that should be taken into account when considering it.

\section{Case study}
\label{Case Study}
In this section, the performance of the four aggregation models is illustrated in the setting of an actual case study. This will provide an insight on how the various models perform, and what difference in the decision-making they induce when applied to real-life data. The case study in question analyses the effects of two treatments (Venlafaxine and Fluoxetine) compared to a placebo, on the effects of treating depression. This study uses data from Nemeroff~\cite{R15}, and expands on the studies conducted by Tervonen \textit{et al.}~\cite{R14} and Saint-Hilary  \textit{et al.}~\cite{R4}. 

Fluoxetine and Venlafaxine are both treatments used to treat depression. Here, the benefit criterion is the treatment response (an increase from baseline score of Hamilton Depression Rating Scale of at least 50\%), and the three risk criteria are nausea, insomnia and anxiety. 

Table~\ref{criteriaCS} shows the outcomes of the trial for the two treatments and the placebo.

\begin{table}[h]
\begin{center}
\noindent\makebox[\textwidth]{
\begin{tabular}{|l   l    l   l|}
\hline
 & Venlafaxine & Fluoxetine  & Placebo   \\
 \hline
Treatment response & 51/96 & 45/100 & 37/101 \\
Nausea  & 40/100 & 22/102 & 8/102  \\
Insomnia  & 22/100 & 15/102 & 14/102  \\
Anxiety & 10/100 & 7/102 & 1/102 \\
\hline
\end{tabular}
}\caption{Number of events and number of patients for each criteria for Venlafaxine, Fluoxetine and Placebo}
\vspace{-2em}
\label{criteriaCS}
\end{center}
\end{table}

For all criteria, we approximate the distributions of the event probabilities by Beta distributions $\mathcal{B}(a,b)$, with $a$ = number of occurrences and $b$ = (number of patients $-$ number of occurrences) of the considered event (response or adverse event), assuming Beta(0,0) priors.  We generated 100,000 samples from each distribution. These samples are then used to approximate the distributions of the linear partial value functions (PVFs) as defined in equation (\ref{eq:partial}) for all criteria and all treatment arms, with the following most and least preferred probabilities of occurrence $\xi'_{j}$ and $\xi''_{j}$:

\begin{itemize}
    \item Most and least preferable values of $\xi'_{j}=0.8$ and $\xi''_{j}=0.2$ for the response,
    \item Most and least preferable values $\xi'_{j}=0$ and $\xi''_{j}=0.5$ for the adverse events.
\end{itemize}

\begin{table}[h]
\begin{center}
\noindent\makebox[\textwidth]{
\begin{tabular}{|lccc|}
\hline
 & Venlafaxine & Fluoxetine  & Placebo   \\
\hline
Treatment response &&& \\
$\xi_{i,1}$& 0.52 (0.42,0.62) & 0.45 (0.35,0.55) & 0.37 (0.28,0.46) \\
$u_{1}(\xi_{i,1}$)& 0.53 (0.37,0.70) & 0.42 (0.26,0.58) & $\mathbf{0.28 (0.13,0.44)}$ \\
Nausea  &&& \\
$\xi_{i,2}$& 0.40 (0.31,0.50) & 0.22 (0.14,0.30) & 0.08 (0.04,0.14)  \\
$u_{2}(\xi_{i,2}$)& $\mathbf{0.20 (0.00,0.39) }$& {0.57 (0.40,0.72) }&{ 0.84 (0.72,0.93)}  \\
Insomnia &&& \\
$\xi_{i,3}$ & 0.22 (0.15,0.31) & 0.15 (0.09,0.22) & 0.14 (0.08,0.21)  \\
$u_{3}(\xi_{i,3}$) & {0.56 (0.39,0.71)} & {0.71 (0.56,0.83) }&{ 0.73 (0.58,0.84) } \\
Anxiety &&&\\
$\xi_{i,4}$& 0.10 (0.05,0.17) & 0.07 (0.03,0.13) & 0.01 (0.00,0.04) \\
$u_{4}(\xi_{i,4}$)&{ {0.80 (0.67,0.90)}} & {0.86 (0.75,0.94)} &{ 0.98 (0.93,1.00)} \\
\hline
\end{tabular}
}\caption{Mean (95$\%$ Credible Interval) of the Beta posterior distributions of benefit and risk parameters and of corresponding PVFs for Venlafaxine, Fluoxetine and Placebo (with values in \textbf{bold} corresponding to those that leading to significant differences between models).  }
\label{PVF}
\end{center}
\end{table}

This case study considers three different weighting combinations, which were used under the linear model by Saint-Hilary \textit{et al.}~\cite{R4}. These sets of weights correspond to three different scenarios of the relative importance of the criteria for the stakeholders. The first scenario reflects the case when all four criteria are equally important. The second scenario corresponds to the benefit criterion having more relative importance than all risk criteria together. The third scenario can be considered as a ``safety first'' scenario, in which each risk criterion has a higher weight than the benefit criterion. As discussed in Section~\ref{Weight Elicitation and Mapping}, the weights of the criteria for the product, multi-linear and SLoS models are obtained by mapping. Note, again, that while the multi-linear model might not exactly induce the same average relative importance of the criteria, the proposed procedure suggests to control the contribution of the interaction terms in the decision at the given level of $c=0.20$, and therefore is used for the sake of simplicity. The mapped weights for each of the three scenarios are presented in Table~\ref{Weighting}.

\begin{table}[h]
\begin{center}
\noindent\makebox[\textwidth]{
\begin{tabular}{|l |llll|llll|llll|}
\hline
 & \multicolumn{4}{l|}{Scenario 1}& \multicolumn{4}{l|}{Scenario 2}& \multicolumn{4}{l|}{Scenario 3}\\
 \hline
Model  & $w_{1}$& $w_{2}$& $w_{3}$& $w_{4}$& $w_{1}$& $w_{2}$& $w_{3}$& $w_{4}$& $w_{1}$& $w_{2}$& $w_{3}$& $w_{4}$\\
  \hline
Linear&         0.25 & 0.25 & 0.25 & 0.25& 0.58 & 0.11 & 0.15 & 0.15 & 0.18 & 0.28 & 0.25 & 0.29 \\
Product &        0.25 & 0.25 & 0.25 & 0.25& 0.58 & 0.11 & 0.15 & 0.15 & 0.18 & 0.28 & 0.25 & 0.29 \\
Multi-Linear &   0.20 & 0.20 & 0.20 & 0.20& 0.53 & 0.06 & 0.10 & 0.10 & 0.13 & 0.23 & 0.20 & 0.24 \\
SLoS&           0.30 & 0.30 & 0.30 & 0.30 & 0.56 & 0.16 & 0.21 & 0.21& 0.24 & 0.33 & 0.30 & 0.34  \\
\hline
\end{tabular}
}\caption{Table of mapped weights for each of the three scenarios.}
\vspace{-2em}
\label{Weighting}
\end{center}
\end{table}

Three pairwise comparisons are made: Venlafaxine against Fluoxetine, Venlafaxine against Placebo, and Fluoxetine against Placebo. We consider that one treatment is recommended over another if the probabilities defined in (\ref{eq:probability}) or (\ref{eq:probability-loss}) are greater than $\psi=0.8$. The probabilities of recommendations under all three scenarios and for each aggregation model are given in Table~\ref{CSresults}.

\begin{table}[h]
\begin{center}
\noindent\makebox[\textwidth]{
\begin{tabular}{|c|ccc|}
\hline
Probability of treatment being	& Venlafaxine over 	& Venlafaxine over 	& Fluoxetine over    \\
recommended as best treatment	& Fluoxetine 		& Placebo 		& Placebo \\
\hline
&\multicolumn{3}{|c|}{Scenario 1} \\
\hline
Linear 			    	& 1.7$\%$  &$<$0.1$\%$ & 7.2$\%$ \\
Product 			    & $1.7\%$  &1.6$\%$	    & 37.0$\%$\\
Multi-Linear  			& 1.7$\%$  &$<$0.1$\%$ 	& 9.1$\%$\\
SLoS 				    & 1.8$\%$  &3.7$\%$ 	& 47.3$\%$  \\
\hline&\multicolumn{3}{|c|}{Scenario 2} \\
\hline
Linear 	    			& 48.0$\%$  &64.7$\%$  & 66.9$\%$ \\
Product 			    & 42.6$\%$  &74.9$\%$  & 80.4$\%$ \\
Multi-Linear  			& 46.3$\%$  &63.0$\%$  & 66.3$\%$\\
SLoS 				    & 36.6$\%$  &72.5$\%$  & 81.4$\%$ \\
\hline
&\multicolumn{3}{|c|}{Scenario 3} \\
\hline
Linear 	    			& 0.6$\%$ 	&0$\%$  	& 2.1$\%$ \\
Product 			    & 0.5$\%$ 	&0.1$\%$		& 18.5$\%$ \\
Multi-Linear  			& 0.6$\%$ 	&0$\%$  	& 3.0$\%$\\
SLoS 				    & 0.6$\%$    &0.6$\%$ 		& 30.1$\%$ \\
\hline
\end{tabular}
}\caption{Probability of treatment being recommended as the best treatment against another for the three pairwise comparison, using each of the four aggregation models, for each of the three weighting scenarios.}
\vspace{-2em}
\label{CSresults}
\end{center}
\end{table}

Under the first scenario with the equal weights for all criteria, the treatment with preferable risk criteria values was more likely to be recommended as the three risk criteria altogether have a greater weight than the one benefit criterion. For the comparison between Venlafaxine and Fluoxetine, the probability that Venlafaxine has better benefit-risk characteristics is around 1.7-1.8$\%$ under all four models. For the comparison between Venlafaxine and the placebo, there is only a minor difference in the probability that Venlafaxine has better benefit-risk characteristics ($<$ 0.1 $\%$ in the linear and multi-linear models, 1.6$\%$ in the product model and 3.7$\%$ in the SLoS model), not enough of a difference to change the recommendation. However, when comparing Fluoxetine to the placebo, a notable difference is observed. Under the linear and multi-linear models, the probability of Fluoxetine having the better benefit-risk characteristics is around 7-10$\%$ (suggesting the placebo is much more preferable), whilst this rises to 37$\%$ under the product model and 47.3$\%$ under the SLoS model (suggesting near-parity of treatments). This occurs due to the penalisation of low benefit criterion values for the placebo, where the 95\% credible interval includes values close to zero (in bold in Table \ref{PVF}). These low values are harshly penalised under the product and SLoS models, as they suggest that the placebo induces no treatment benefit with a non-neglectable probability. The linear model does not account for this and strongly favours the placebo, while the multi-linear does not penalise these values strongly. 

Under the second scenario, the treatment response is considered as the most important factor, and is given a weighting greater than that of the three risk criteria combined. For the comparison between Venlafaxine and Fluoxetine, both the product and SLoS models say that Venlafaxine has inferior benefit-risk characteristics (42.6$\%$ and 36.6$\%$ probability of being better, respectively). More average results are observed with both the linear model, which gives a probability of 48.0$\%$, and the multi-linear model, which gives a probability of 46.3$\%$. Again, the difference between the probability of the linear model and those of the product and SLoS models is due to the penalising effects of the latter. This occurs because of the nausea risk criterion interval contains zero for Venlafaxine (in bold in Table \ref{PVF}), which causes the product and SLoS models to recommend Fluoxetine more often than  Venlafaxine, despite the weighing criteria giving preference to the treatment response (which is greater with Venlafaxine). With the multi-linear model, the penalisation of the undesirable nausea criterion is not as strong as in the product or SLoS models, as the weight mapping induces a drop from 0.11 to 0.06 in the weight given to the corresponding individual term, and the effect of the interaction terms is not enough to overcome this. 

For the comparison between Venlafaxine and the placebo, the probability that Venlafaxine has better benefit-risk characteristics is between 63-75$\%$ across the four models. The product and SLoS models both penalise the low benefit value of the placebo, which is why they are both more likely to recommend Venlafaxine than the other two models. Additionally, the product and SLoS models both also penalise the nausea criterion value of Venlafaxine, and due to the increase weighting given to it by the SLoS model mapping, this causes the product model to be more likely to recommend Venlafaxine than the SLoS model.

For the comparison between Fluoxetine and the placebo, the probability that Fluoxetine has better benefit-risk characteristics is around 65-80$\%$ under all four models, with the probability of Fluoxetine being preferable increasing as the methods increase the penalisation applied to the placebo's lack of benefit effect. The stronger penalisation occurs under the product and SLoS models, hence why they are both more likely to recommend Fluoxetine. 

Across all three comparisons, the multi-linear model is always slightly less likely to recommend the treatment with the greater benefit value than the linear model. As this is the scenario where the benefit criterion is considered to be the most important, this shows that the weight splitting with the multi-linear model induces a loss of the preferences that were given when the weights were originally set out for the linear model, illustrating some of the problems theorised in the methods section.

Under the third scenario, a ``safety first" approach is adopted, giving the risk factors a higher weighting.
The probability that Venlafaxine has better benefit-risk characteristics is around 0.5-0.6$\%$ when it is compared to Fluoxetine and around 0-0.6$\%$ when it is compared to placebo, under all four models. For the comparison between Fluoxetine and the placebo, the probability that Fluoxetine has better benefit-risk characteristics is around 2.1-3.0$\%$ for the linear and multi-linear models, whilst this increases to 18.5$\%$ under the product model and 30.1$\%$ under the SLoS model. This increase occurs for the same reasons outlined for the same comparison in scenario 1: The penalisation of the benefit criterion for the placebo, with its 95\% credible interval including low values (in bold in Table \ref{PVF}). The linear model does not account for this and strongly favours the placebo, while the multi-linear does not penalise these values sufficiently and still favours the placebo.

Overall, this case study provides us with a number of important observations shedding a light on the differences in the aggregation performances. Firstly, the effects of extremely undesirable outcomes (those highlighted in bold in Table \ref{PVF}) are more significantly and consistently penalised in the product and SLoS models (the penalisation is stronger in the SLoS model than the product model, although they give the same recommendation for every comparison). 
These examples also help to show that the models provide similar recommendations when one treatment is clearly preferable than its competitor.
Lastly, the weight splitting in the multi-linear model induces a change in the relative importance between criteria that may not always reflect the choices of weights as well as other models, highlighted in scenario 2. This makes it less appealing than other models. 

To draw further conclusions regarding the differences between models, we conduct a comprehensive simulation study under various scenarios and under their many different realisations.

\section{Simulation Study}
\label{Simulation Study}
\subsection{Setting}
\label{Setting}
\begin{figure}[h]
\begin{center}
\includegraphics[width=16cm]{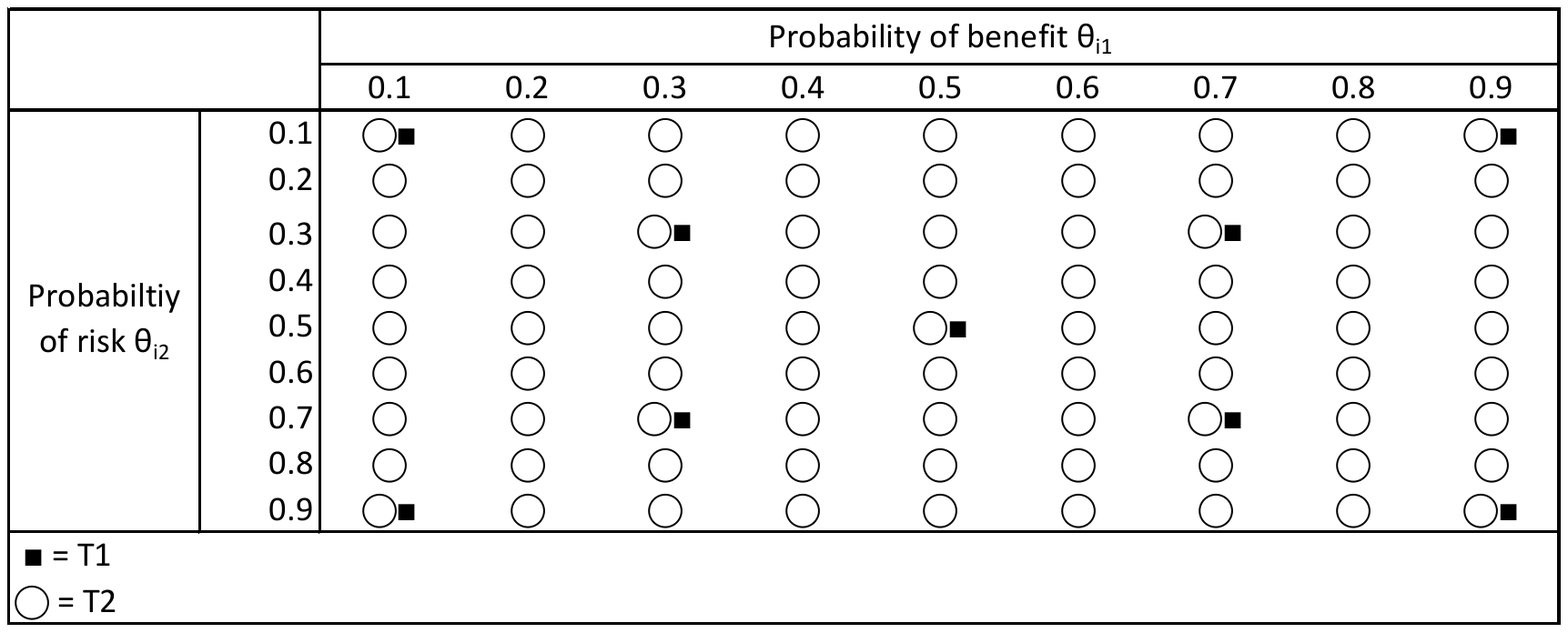}
\end{center}
\vspace{-38em}
\caption{Simulation scenarios for the trial with two criteria.}
\label{fig:tg}
\end{figure}

To evaluate the performances of the four aggregation models, a comprehensive simulation study covering a wide range of possible clinical cases is conducted. This allows us to investigate many scenarios and their various realisations rather than a single dataset as in the case study. The simulation study is preformed in a setting with two treatments, named $T_1$ and $T_2$, that are compared in randomised clinical trials with $N=100$ patients allocated to each treatment. Each treatment is evaluated based on two criteria: one benefit ($j=1$) and one risk ($j=2$). We assume that benefit events are desirable (e.g. treatment response), while risk events should be avoided (e.g. adverse event), with $\theta_{ij}$ being their true probability of occurrence for each treatment  $i=1, 2$. The PVFs are defined as $u_1(\theta_{i1})=\theta_{i1}$ and $u_2(\theta_{i2})=1-\theta_{i2}$. The two criteria are deemed equally important and therefore are given equal weighting criteria. We start with the case of uncorrelated criteria and explore the effect of the presence of correlations in Section~\ref{Sensitivity Analysis: Correlation in criteria}.
The range of true values of the benefit and risk criteria and the corresponding simulation scenarios are given in Figure~\ref{fig:tg}.

Figure~\ref{fig:tg} shows all the different values that the benefit and risk criteria can take for both $T_1$ and $T_2$, where black squares correspond to the pairs of criterion values for $T_1$ and white circles correspond to the pairs of criterion values for $T_2$. For each of the nine fixed characteristics of $T_1$, all 81 possible values of $T_2$, with $\theta_{2,1},\theta_{2,2} \in (0.1 ,0.2,\ldots, 0.9)$ are considered, resulting in 729 scenarios. The fixed characteristics for $T_1$ are referred to as follows:\\
Scenario 1: $T_1$=($\theta_{1,1}$=0.5,$\theta_{1,2}$=0.5) ~~~~~~~~~~~
Scenario 2: $T_1$=($\theta_{1,1}$=0.3,$\theta_{1,2}$=0.7) \\
Scenario 3: $T_1$=($\theta_{1,1}$=0.7,$\theta_{1,2}$=0.3) ~~~~~~~~~~~
Scenario 4: $T_1$=($\theta_{1,1}$=0.1,$\theta_{1,2}$=0.1)\\
Scenario 5: $T_1$=($\theta_{1,1}$=0.9,$\theta_{1,2}$=0.9) ~~~~~~~~~~~
Scenario 6: $T_1$=($\theta_{1,1}$=0.3,$\theta_{1,2}$=0.3)\\
Scenario 7: $T_1$=($\theta_{1,1}$=0.7,$\theta_{1,2}$=0.7) ~~~~~~~~~~~
Scenario 8: $T_1$=($\theta_{1,1}$=0.9,$\theta_{1,2}$=0.1) \\
Scenario 9: $T_1$=($\theta_{1,1}$=0.1,$\theta_{1,2}$=0.9)) ~~~~~~~~~~~
where $\theta_{1,j}$ is the true value of criterion $j$ for $T_{1}$. 

\subsection{Data Generation and Comparison Procedure}
\label{Data Generation and Comparison Procedure}
The following Bayesian procedure is used for the simulation study:
\begin{itemize}
    \item Step 1: Simulate randomised clinical trials with two treatments $T_{1}$ and $T_{2}$, each with two uncorrelated criteria, and the sample size of $N=100$ in each treatment arm. 
    \item Step 2: Derive the posterior distributions using the simulated data assuming a degenerate prior, Beta(0,0), to reduce the influence of the prior distribution. Draw 2000 samples from each posterior distribution of the criteria and obtain the corresponding empirical distribution for the PVF. 
    
    \item Step 3: Use the posterior distributions of the PVF in each of the aggregation models as given in Equations \ref{eq:difference} and \ref{eq:difference-loss} to compute the probability in Equations~\ref{eq:probability} and \ref{eq:probability-loss} that treatment $T_1$ has better benefit-risk profile, $\mathcal{P}_{X}^{1,2}$ (for some model $X$), and compare to the threshold value $\psi=0.8$. If $\mathcal{P}_{X}^{1,2}>0.8$, then treatment $T_1$ is recommended. If $\mathcal{P}_{X}^{1,2}<0.2$, then treatment $T_2$ is recommended. If 0.2 $\leq \mathcal{P}_{X}^{1,2} \leq 0.8$, then neither treatment is recommended.
    \item Step 4: Repeat steps 1-3 for 2500 simulations trials.
    \item Step 5: Estimate the probability that each treatment is recommended $\big(\mathbb{P} \big(\mathcal{P}_{X}^{1,2} > 0.8\big)\big)$ by its proportion over 2500 simulated trials. 
    \end{itemize}

The aggregation models will be compared using $\mathbb{P}\big(\mathcal{P}_{X}^{1,2}>0.8 \big)$, which is the probability that the model $X$ recommends $T_{1}$ over $T_{2}$, and $\phi_{X-Y}=\mathbb{P}\big(\mathcal{P}_{X}^{1,2}>0.8 \big)-\mathbb{P}\big(\mathcal{P}_{Y}^{1,2}>0.8 \big)$, which is the difference between the probability that the model $X$ recommends $T_{1}$ and the probability that the model $Y$ recommends $T_{1}$. The value of $\phi$ represents a difference between two probabilities, and can therefore take the range of values $-1 \leq \phi \leq 1$. If $0 < \phi \leq 1$, then the model $X$ recommends $T_{1}$ more often than model $Y$. If $-1 \leq \phi < 0$, then the model $Y$ recommends $T_{1}$ more often than model $X$. If $\phi=0$, then the two models make the recommendations with the same probability. Note that, for the ML model, we adopt $c=0.20$ as in the case study above.

\subsection{Results}
\label{Results}
\begin{figure}[h]
\begin{center}
\includegraphics[width=17cm]{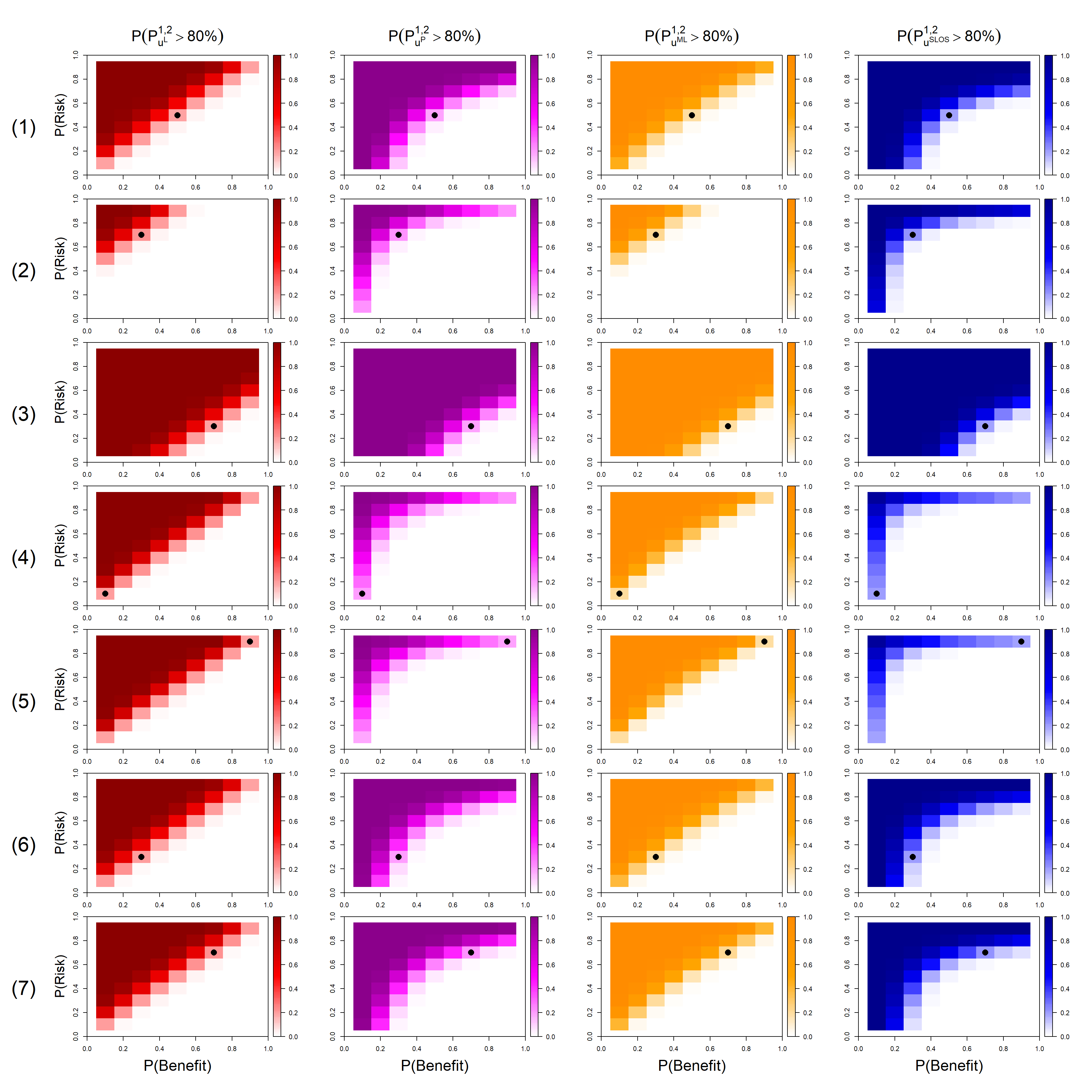}
\end{center}

\caption{Probability that the model recommends $T_{1}$ over $T_{2}$, $\mathbb{P}\big(\mathcal{P}^{1,2}>0.8$\big), for scenarios 1-7 for the linear (red), product (purple), multi-linear (orange), and SLoS (blue) models. }

\label{fig:sim1}
\end{figure}

\begin{figure}[h]
\begin{center}
\includegraphics[width=17cm]{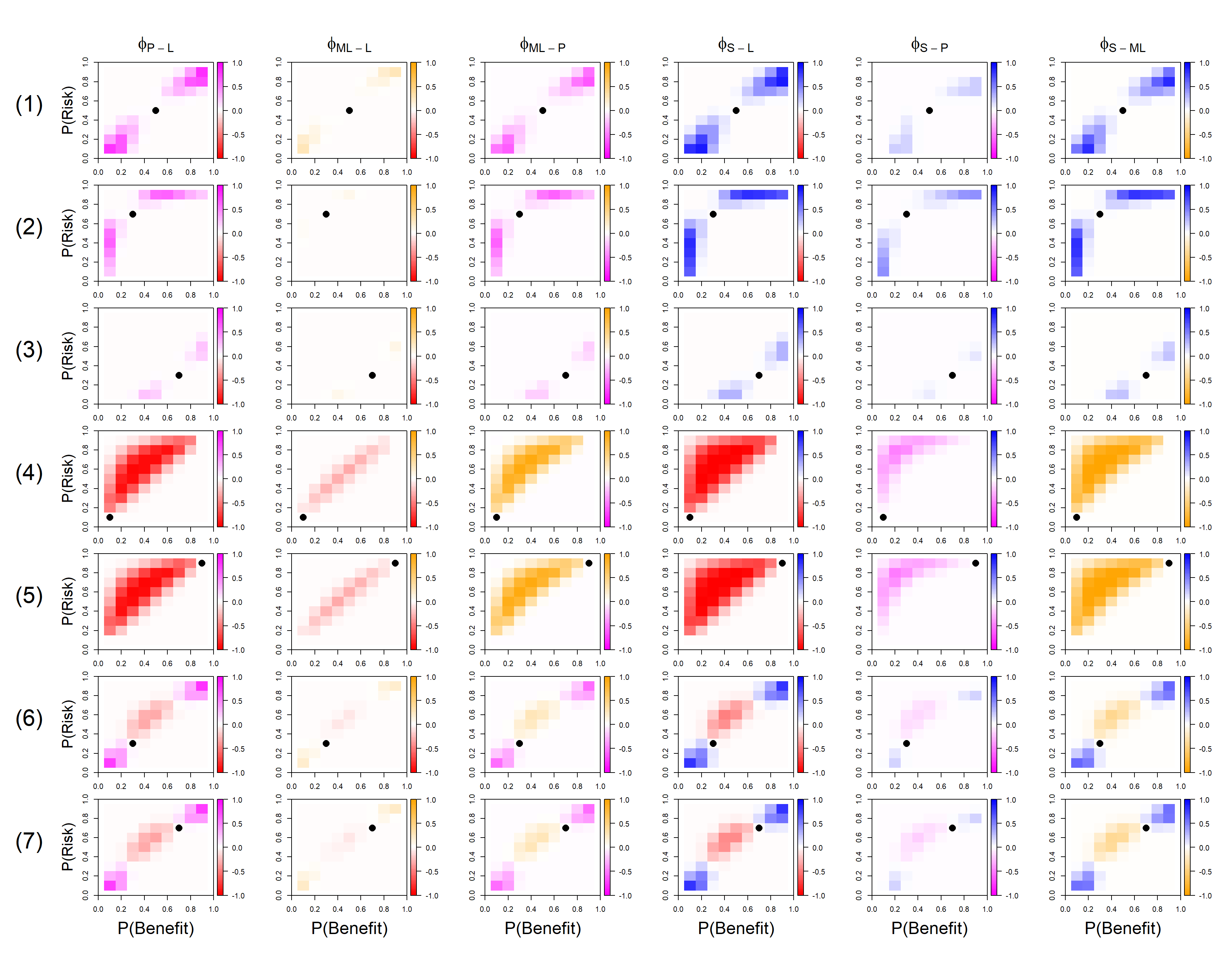}
\end{center}
\vspace{-2em}
\caption{Results of the six pairwise comparisons of the four AM, where a cell being a colour indicates that that AM recommended $T_{1}$ more than the comparative AM (the deeper the colour, the greater the difference in recommendation).}
\vspace{-2em}
\label{fig:sim2}
\end{figure}

The results are presented on Figures \ref{fig:sim1} and \ref{fig:sim2}. The first seven scenarios referred to above for treatment $T_{1}$ are presented in the rows labeled 1-7. Each graph corresponds to fixed expected probabilities of event for treatment $T_{1}$, and each cell corresponds to a combination of expected probabilities of benefit and risk for $T_{2}$. When reference is made to the ``diagonal'', this refers to the diagonal line that runs from the bottom left corner of the graph to the top right. In all scenarios, all models agree to recommend $T_{1}$ when it is undoubtedly better than $T_{2}$ i.e. when $T_{1}$ is more effective and less harmful than $T_{2}$ (or to recommend $T_{2}$ when $T_{1}$ is indisputably worse, i.e. less effective and more toxic). For this reason, the results for scenarios 8 and 9 are not presented here, but are included in the Supplementary Material for completeness.

The probabilities $\mathbb{P}\big(\mathcal{P}_{L}^{1,2}>0.8$\big) (Red),~ $\mathbb{P}\big(\mathcal{P}_{P}^{1,2}>0.8$\big) (Purple),~ $\mathbb{P}\big(\mathcal{P}_{ML}^{1,2}>0.8$\big) (Orange), and $\mathbb{P}\big(\mathcal{P}_{S}^{1,2}>0.8$\big) (Blue) are shown in Figure \ref{fig:sim1}, and all six pairwise comparisons in these probabilities are given in Figure~\ref{fig:sim2}. From left to right, Figure \ref{fig:sim2} shows $\phi_{P-L},\phi_{ML-L},\phi_{ML-P},\phi_{S-L},\phi_{S-P}$ and $\phi_{S-ML}$.

In Figure \ref{fig:sim2}, a colour of a cell corresponds to the aggregation model of this colour to recommend treatment $T_1$ with higher probability than another method. For instance, red cells in the first column of Figure \ref{fig:sim2} showing ($\phi_{P-L}$) indicate that, when $T_{2}$ characteristics take the corresponding value, the linear model recommends $T_{1}$ more often than the product one.

In Scenario 1, the four models are in agreement to recommend $T_{1}$ when $T_{2}$ corresponds to less benefit and more risk. On the diagonal, the product and SLoS models both favour $T_{1}$ over $T_{2}$ when $T_{2}$ has either extremely high benefit and risk (top right corner), or extremely low benefit and risk (left bottom corner), compared to either the linear or multi-linear models. This occurs due to the penalisation of extremely low benefit and extremely high risk by the product and SLoS models. Comparing product and SLoS models for these values of benefit-risk, SLoS favours $T_{1}$ over $T_{2}$ more often for low but not boundary values of the criteria. This occurs due to the SLoS model penalising the undesirable qualities more than the product one (this is similar to trends observed in the case study). Compared to the linear model, the multi-linear model recommends $T_{1}$ over $T_{2}$ with higher probability when $T_{2}$ has either higher benefit and higher risk, or lower benefit and lower risk due to the interaction term providing mild penalisation of extremely high risk or extremely low benefit.
However, there is (in most cases), a greater magnitude of difference between the SLoS and product models than between the linear and multi-linear models.

For example, when $T_{2}$ has criteria values $\theta_{2,1}=0.2$ (benefit), $\theta_{2,2}=0.1$ (risk) (lower benefit, lower risk), $T_{1}$ is recommended in 2$\%$ of the trials under the linear model, in 70$\%$ under the product, in 8$\%$ under the multi-linear and in 90$\%$ under SLoS. 
 This tells us that the product and SLoS models do not permit that the decrease in risk is worth the decrease in benefit that comes with it (the SLoS model more than the product model), whilst the linear and multi-linear models both consider it acceptable. Considering the case when $\theta_{2,1}=0.7$, $\theta_{2,2}=0.7$ (higher benefit and higher risk risk compared to $T_1$), $T_{1}$ is recommended in 20$\%$ of the trials under the linear model compared to 49$\%$ for product model, 25$\%$ for the multi-linear model and 61$\%$ for SLoS model. 
This tells us that the product and SLoS models do not permit that the increase in benefit is worth the increase in risk that comes with it (again, this effect is stronger in the SLoS model than the product model), whilst the linear and multi-linear both consider it acceptable (the linear model more-so than the multi-linear model). Similar observations can be made in Scenarios 2-3.

However, a distinguishing difference between the designs under Scenario 1 can be found when $T_{2}$ has the criteria $\theta_{2,1}=0.9$, $\theta_{2,2}=0.7$. In this comparison, $T_{1}$ is recommended in 0$\%$ of the trials under the linear model compared to 11$\%$ for product model, 0$\%$ for the multi-linear model and 30$\%$ for SLoS model. Meanwhile, $T_{2}$ is recommended in 92$\%$ of the trials under the linear model compared to 32$\%$ for product model, 84$\%$ for the
multi-linear model and 13$\%$ for SLoS model. This shows that the linear, product and multi-linear models are all more likely to recommend $T_{2}$, whilst only the SLoS model is more likely to recommend $T_{1}$. This occurs due to the different strengths of penalisation between the models, and only the SLoS model does not consider this an acceptable trade-off. This shows that the product model and the SLoS model do not always make the same recommendations, and that these differences can sometimes be quite large. 

In Scenario 4, where $T_{1}$ has extremely low benefit and risk, it is very rarely recommended by either the product of SLoS models, whereas it recommended by both the linear and multi-linear models, in cases where $T_{2}$ has some increase in benefit, but a higher increase in risk. This occurs because the SLoS and product models penalise extremely low benefit so severely that the level of risk has almost no impact on the recommendation. The multi-linear model also penalises the extreme low benefit, but on a much smaller scale.
For example, for $T_{2}$ with criteria values $\theta_{2,1}=0.6$, $\theta_{2,2}=0.7$, $T_{1}$ is recommended with probability 68$\%$ under the linear model, never recommended under the product model, 41$\%$ under the multi-linear model and never recommended under the SLoS model. This shows that the product and SLoS models reflect the desirable properties outlined above: that we are not interested in the risk criterion value of a treatment if the benefit criterion value is small/zero, whilst both the linear and multi-linear models do not reflect this (although the multi-linear model does somewhat penalise this).
Similar results are observed in Scenario 5, where $T_{1}$ has extreme risk and extreme benefit. The SLoS and product models will recommend $T_{2}$ if it has lower risk than $T_{1}$ as long as it has some benefit, whereas the linear model and the multi-linear model will recommend $T_{1}$ over $T_{2}$ if the benefit of $T_{2}$ decreases by a greater amount than the risk.

It should be noted that poor recommendations can be made under the product and SLoS models if both $T_{1}$ and $T_{2}$ have a risk criterion value of 0.9, as the strength of the penalisation of the undesirable criteria overpowers the effect of the benefit. For example, in scenario 5 where $T_{2}$ has criteria values $\theta_{2,1}=0.8$, $\theta_{2,2}=0.9$ (same risk criterion value as $T_{1}$ but a lower benefit criterion value), $T_{1}$ is recommended with probability 75$\%$ under the linear model, 27$\%$ under the product model, 68$\%$ under the multi-linear model and 23$\%$ under the SLoS model (this effect is stronger in the SLoS model than in the product model due to its harsher penalisation of the undesirable criteria). They both did recommend $T_{2}$ with probabilities 13$\%$ and 17$\%$ respectively, showing that they still recommend the better treatment $T_{1}$ more often than $T_{2}$, but that these two models hardly discriminate very unsafe drugs (for comparison, both the linear and multi-linear models only recommended $T_{2}$ with probability 1$\%$ each). 

In Scenarios 6-7, all AM recommend $T_{1}$ over $T_{2}$ when $T_{2}$ is unarguably worse (similarly they all recommend $T_{2}$ over $T_{1}$ when $T_{1}$ is unarguably worse). Along the diagonal, the SLoS model recommends $T_{1}$ over $T_{2}$ more often than the other AM when $T_{2}$ has either extreme low benefit and extreme low risk, or extreme high benefit and extreme high risk, compared to $T_{1}$ (although the product model recommends $T_{1}$ only a slightly smaller proportion of times than the SLoS model). Again, this is the result of the penalisation of extremely low benefit or extremely high risk criteria. Similarly, the multi-linear model recommends $T_{1}$ over $T_{2}$ more often than the linear model in the same circumstances. For example, in Scenario 6, when $T_{2}$ has criteria values $\theta_{2,1}=0.2$, $\theta_{2,2}=0.2$ (lower benefit and lower risk), $T_{1}$ is recommended with probability 21$\%$ under the linear model, 59$\%$ under the product model, 28$\%$ under the multi-linear model and 68$\%$ under the SLoS model. This shows how the different levels of penalisation affect the recommendations, where the stronger the penalisation of the undesirable low benefit criterion value, the more likely an AM is to recommend $T_{1}$, and is the reason why there is such a large difference between the linear and SLoS models recommendations.

Overall, the simulation study has shown that, for the two criteria having an equal relative importance, SLoS penalises extremely low benefit and extremely high risk criteria the most, whilst the product model penalises these moderately, acting as a sort of middle ground between the linear and SLoS models. The multi-linear model offers a small amount of penalisation (less than the product model), but due to the added complexity of this model when more criteria are added, it should not be recommended over either the SLoS model or the product model. The linear and multi-linear models both recommend treatments with no benefit/high risk over other viable alternatives, which contradicts conditions set out by Saint-Hilary \textit{et al.}~\cite{R6}. Therefore we can provisionally conclude that the two models that appeal most at this point are the product and SLoS models.

\subsection{Sensitivity Analysis: Correlated criteria}
\label{Sensitivity Analysis: Correlation in criteria}
The results above concerned the case with the two criteria being uncorrelated. However, it might be reasonable to assume that the criteria for one treatment might be correlated. In this section, we study how robust the recommendation by each of the four models are to the correlation between the benefit and risk criteria. We consider two cases of the correlation: a strong positive correlation ($\rho=0.8$) and a strong negative correlation ($\rho=-0.8$) between the criteria. The correlated outcomes were generated using a procedure laid out in \citet{R17}

We study how likely the correlated outcomes are to change the final recommendation of one of the treatments. Specifically, we study the proportion of cases under each of the scenarios in which the difference in the probability of recommending treatment $T_1$, $\mathbb{P} \big(\mathcal{P}_{X}^{1,2} > 0.8\big)$, changes by more than 2.5\% and by 5\%. Table \ref{fig:pcorr} show the number of cases (out of 81) under each of nine scenarios, in which the differences in the probabilities to recommend $T_{1}$ over $T_{2}$ changes by at least 2.5\% and 5\% comparing the positively correlated and uncorrelated criteria. The case investigating the effects of negative correlation shows similar results to those presented here, and is included in the Supplementary Material. For example, the first entry in Table \ref{fig:pcorr} shows that in 37\% cases under Scenario 1, the probability to recommend $T_{1}$ changes by at least 2.5\% if the linear model is used.

\begin{table}[h]
\begin{center}
\noindent\makebox[\textwidth]{
\begin{tabular}{|cc|cccc|}
\hline
    & &Linear Model & Product Model & Multi-Linear Model & SLoS Model \\
\hline
Scenario 1    &     \begin{tabular}{@{}c@{}}$\geq$2.5$\%$ \\ $\geq$5$\%$\end{tabular} 
                & \begin{tabular}{@{}c@{}}30/81 (37.0$\%$) \\ 22/81 (27.2$\%$)\end{tabular} 
                & \begin{tabular}{@{}c@{}}24/81 (29.6$\%$) \\ 15/81 (18.5$\%$)\end{tabular} 
                & \begin{tabular}{@{}c@{}}29/81 (35.8$\%$) \\ 21/81 (25.9$\%$)\end{tabular} 
                & \begin{tabular}{@{}c@{}}22/81 (27.2$\%$)\\ 12/81 (14.8$\%$)\end{tabular} \\
\hline                
Scenario 2    &     \begin{tabular}{@{}c@{}}$\geq$2.5$\%$ \\ $\geq$5$\%$\end{tabular} 
                & \begin{tabular}{@{}c@{}}10/81 (12.3$\%$) \\ 5/81 (6.2$\%$)\end{tabular} 
                & \begin{tabular}{@{}c@{}}15/81 (18.5$\%$)\\ 3/81 (3.7$\%$)\end{tabular} 
                & \begin{tabular}{@{}c@{}}15/81 (18.5$\%$)\\ 5/81 (6.2$\%$)\end{tabular} 
                & \begin{tabular}{@{}c@{}}12/81 (14.8$\%$)\\ 3/81 (3.7$\%$)\end{tabular} \\
\hline
Scenario 3    &     \begin{tabular}{@{}c@{}}$\geq$2.5$\%$ \\ $\geq$5$\%$\end{tabular} 
                & \begin{tabular}{@{}c@{}}16/81 (19.8$\%$)\\ 6/81 (7.4$\%$)\end{tabular} 
                & \begin{tabular}{@{}c@{}}13/81 (16.0$\%$)\\ 5/81 (6.2$\%$)\end{tabular} 
                & \begin{tabular}{@{}c@{}}17/81 (21.0$\%$)\\ 4/81 (4.9$\%$)\end{tabular} 
                & \begin{tabular}{@{}c@{}}14/81 (17.3$\%$)\\ 4/81 (4.9$\%$)\end{tabular} \\
\hline
Scenario 4    &     \begin{tabular}{@{}c@{}}$\geq$2.5$\%$ \\ $\geq$5$\%$\end{tabular} 
                & \begin{tabular}{@{}c@{}}22/81 (27.2$\%$)\\ 17/81 (21.0$\%$)\end{tabular} 
                & \begin{tabular}{@{}c@{}}3/81  (3.7$\%$)\\ 0/81 (0$\%$)\end{tabular} 
                & \begin{tabular}{@{}c@{}}22/81 (27.2$\%$)\\ 15/81 (18.5$\%$)\end{tabular} 
                & \begin{tabular}{@{}c@{}}0/81  (0$\%$)\\ 0/81 (0$\%$)\end{tabular} \\
\hline
Scenario 5    &     \begin{tabular}{@{}c@{}}$\geq$2.5$\%$ \\ $\geq$5$\%$\end{tabular} 
                & \begin{tabular}{@{}c@{}}23/81 (28.4$\%$)\\ 17/81 (21.0$\%$)\end{tabular} 
                & \begin{tabular}{@{}c@{}}4/81  (4.9$\%$)\\ 0/81 (0$\%$)\end{tabular} 
                & \begin{tabular}{@{}c@{}}22/81 (27.2$\%$)\\ 15/81 (18.5$\%$)\end{tabular} 
                & \begin{tabular}{@{}c@{}}0/81  (0$\%$)\\ 0/81 (0$\%$)\end{tabular} \\
\hline
Scenario 6    &     \begin{tabular}{@{}c@{}}$\geq$2.5$\%$ \\ $\geq$5$\%$\end{tabular} 
                & \begin{tabular}{@{}c@{}}29/81 (35.8$\%$)\\ 20/81 (24.7$\%$)\end{tabular} 
                & \begin{tabular}{@{}c@{}}21/81 (25.9$\%$)\\ 12/81 (14.8$\%$)\end{tabular} 
                & \begin{tabular}{@{}c@{}}30/81 (37.0$\%$)\\ 16/81 (19.8$\%$)\end{tabular} 
                & \begin{tabular}{@{}c@{}}17/81 (21.0$\%$)\\ 4/81 (4.9$\%$)\end{tabular} \\
\hline
Scenario 7    &     \begin{tabular}{@{}c@{}}$\geq$2.5$\%$ \\ $\geq$5$\%$\end{tabular} 
                & \begin{tabular}{@{}c@{}}25/81 (30.9$\%$)\\ 14/81 (17.3$\%$)\end{tabular} 
                & \begin{tabular}{@{}c@{}}19/81 (23.5$\%$)\\ 9/81 (11.1$\%$)\end{tabular} 
                & \begin{tabular}{@{}c@{}}24/81 (29.6$\%$)\\ 15/81 (18.5$\%$)\end{tabular} 
                & \begin{tabular}{@{}c@{}}13/81 (16.0$\%$)\\ 2/81 (2.5$\%$)\end{tabular} \\
\hline
Scenario 8    &     \begin{tabular}{@{}c@{}}$\geq$2.5$\%$ \\ $\geq$5$\%$\end{tabular} 
                & \begin{tabular}{@{}c@{}}0/81 (0$\%$)\\ 0/81 (0$\%$)\end{tabular}  
                & \begin{tabular}{@{}c@{}}0/81 (0$\%$)\\ 0/81 (0$\%$)\end{tabular}  
                & \begin{tabular}{@{}c@{}}0/81 (0$\%$)\\ 0/81 (0$\%$)\end{tabular} 
                & \begin{tabular}{@{}c@{}}0/81 (0$\%$)\\ 0/81 (0$\%$)\end{tabular}  \\
\hline
Scenario 9    &     \begin{tabular}{@{}c@{}}$\geq$2.5$\%$ \\ $\geq$5$\%$\end{tabular} 
                & \begin{tabular}{@{}c@{}}1/81 (1.2$\%$)\\ 0/81 (0$\%$)\end{tabular} 
                & \begin{tabular}{@{}c@{}}1/81 (1.2$\%$)\\ 0/81 (0$\%$)\end{tabular} 
                & \begin{tabular}{@{}c@{}}1/81 (1.2$\%$)\\ 0/81 (0$\%$)\end{tabular} 
                & \begin{tabular}{@{}c@{}}1/81 (1.2$\%$)\\ 0/81 (0$\%$)\end{tabular} \\

\hline
Total&   \begin{tabular}{@{}c@{}}$\geq$2.5$\%$ \\ $\geq$5$\%$\end{tabular} 
                & \begin{tabular}{@{}c@{}}156/729 (21.4$\%$)\\ 101/729 (13.9$\%$)\end{tabular} 
                & \begin{tabular}{@{}c@{}}100/729 (13.7$\%$)\\ 44/729 (6.0$\%$)\end{tabular} 
                & \begin{tabular}{@{}c@{}}160/729 (21.9$\%$)\\ 91/729 (12.5$\%$)\end{tabular} 
                & \begin{tabular}{@{}c@{}}79/729  (10.88$\%$)\\ 25/729 (3.4$\%$)\end{tabular}  \\
\hline
\end{tabular}
}\caption{Number of times ($\%$) when the difference in recommending $T_{1}$ changes by at least 2.5$\%$ or 5$\%$ between the positively correlated criteria and the non-correlated criteria}
\vspace{-2em}
\label{fig:pcorr}
\end{center}
\end{table}

Table \ref{fig:pcorr} shows that all four models are the most affected by correlation under Scenario 1 with the characteristics of $T_1$ being in the middle of the unit interval. This effect is, however, less prominent for the Product and SLoS Models. At the same time, under Scenarios 2-7, the correlation has a larger effect on the linear and multi-linear models than on the other two models. Scenarios 8-9 are hardly affected by any correlation, and the effect is similar across all four models. 

Overall, the SLoS model is the least affected by correlation between the criteria, the product model is the second least affected whereas the multi-linear (for the threshold 2.5\%) and the linear model (for the threshold 5\%) are the most affected ones.

\section{Discussion}
\label{Discussion}

In this article, four potential AM are investigated for use in benefit-risk analyses: The linear model, product model, multi-linear model and the SLoS model. The differences of these models were highlighted in a case-study and a simulation study.

In most clear cases (i.e. when one treatment has more benefit and less risk than the competitor), all AM gave similar recommendations. However, in cases where one treatment had either no benefit or extreme risk, the models which penalised undesirable values more (the product and SLoS models) gave more desirable recommendations: non-effective or extremely unsafe treatments are never recommended. Furthermore, with these models, more risk is accepted in order to increase benefit when the amount of benefit is small than when it is high (or less benefit is desirable to reduce risk when the amount of risk is high than when it is small), which is consistent with the well established assumption of non-linearity of human preferences~\cite{R7}. It should be noted that these models hardly discriminate two treatments that slightly differ but have both extremely undesirable properties. However, in this case, none of the treatments should be recommended anyway.

The effects of correlations between criteria was also investigated in this study. The overall effect of correlations was small to negligible in the product and SLoS models, showing these AM are not much affected by correlations between the criteria. However, the linear and multi-linear models were more likely to see a 2.5$\%$ or 5$\%$ change in the probability of recommending one treatment over another, showing that they are more affected by correlations between the criteria.

Overall, the two models to recommend from this investigation are the product model and the SLoS model, depending on how severely the decision-maker whish to penalise treatments with either no benefit or extreme risk (moderate penalisation: product model, strong penalisation: SLoS model). The multi-linear model, whilst acting as a middle ground between the linear model and the product and SLoS models in the simulation study, involves an increased complexity behind the model. These include the increased complexity involved with adding additional terms and increased difficulty in weight mapping. This model also struggled to truly reflect the weightings given in the case study, especially in scenario 2. Because of this, we do not recommend this AM over the product or SLoS models. Additionally, the linear and multi-linear models should not be recommended as both of these models do not contain the two desirable properties outlined in Saint-Hilary \textit{et al.}\cite{R6}: That treatments with no benefit/extreme risk should not be recommended, and that a larger increase in risk is accepted in order to increase the benefit if the benefit is small compared to if the benefit is high -- both of which are present in the product and SLoS models.

\section*{Supplemental Material}
\label{Supplemental material}
Supplemental material available at: https://github.com/Tom-Menzies/Code-Menzies-2020

\section*{Acknowledgements}
This report is independent research supported by the National Institute for Health Research (NIHR Advanced Fellowship, Dr Pavel Mozgunov,
NIHR300576). The views expressed in this publication are those of the
authors and not necessarily those of the NHS, the National Institute for Health Research or the Department of Health and Social Care (DHCS). 

\bibstyle{pe}

\end{document}


\runninghead{Menzies, Saint-Hilary and Mozgunov}

\title{Supplemental Material - A Comparison of Various Aggregation Functions in Multi-Criteria Decision Analysis for Drug Benefit-Risk Assessment}

\author{Tom Menzies\affilnum{1,2}, Gaelle Saint-Hilary\affilnum{3,4} and Pavel Mozgunov\affilnum{5}}

\affiliation{\affilnum{1}Clinical Trials Research Unit, Leeds Institute of Clinical Trials Research, University of Leeds, Leeds, UK \\
\affilnum{2}Department of Mathematics and Statistics, Lancaster University, Lancaster, UK  \\
\affilnum{3}Department of Biostatistics, Institut de Recherches Internationales Servier (IRIS), Suresnes, France\\
\affilnum{4}Dipartimento di Scienze Matematiche (DISMA) Giuseppe Luigi Lagrange, Politecnico di Torino, Torino, Italy\\
\affilnum{5}Medical and Pharmaceutical Statistics Research Unit, Department of Mathematics and Statistics, Lancaster University, Lancaster, UK }

\corrauth{Tom Menzies, Clinical Trials Research Unit,
University of Leeds,
Leeds,
LS2 9JT, UK.}
\email{t.menzies@leeds.ac.uk}

\maketitle

\setcounter{figure}{0}
\renewcommand{\thefigure}{S\arabic{figure}}

\setcounter{table}{0}
\renewcommand{\thetable}{S\arabic{table}}

\clearpage

\section{Additional simulation results}
\begin{figure}[h]
\begin{center}
\includegraphics[width=17cm]{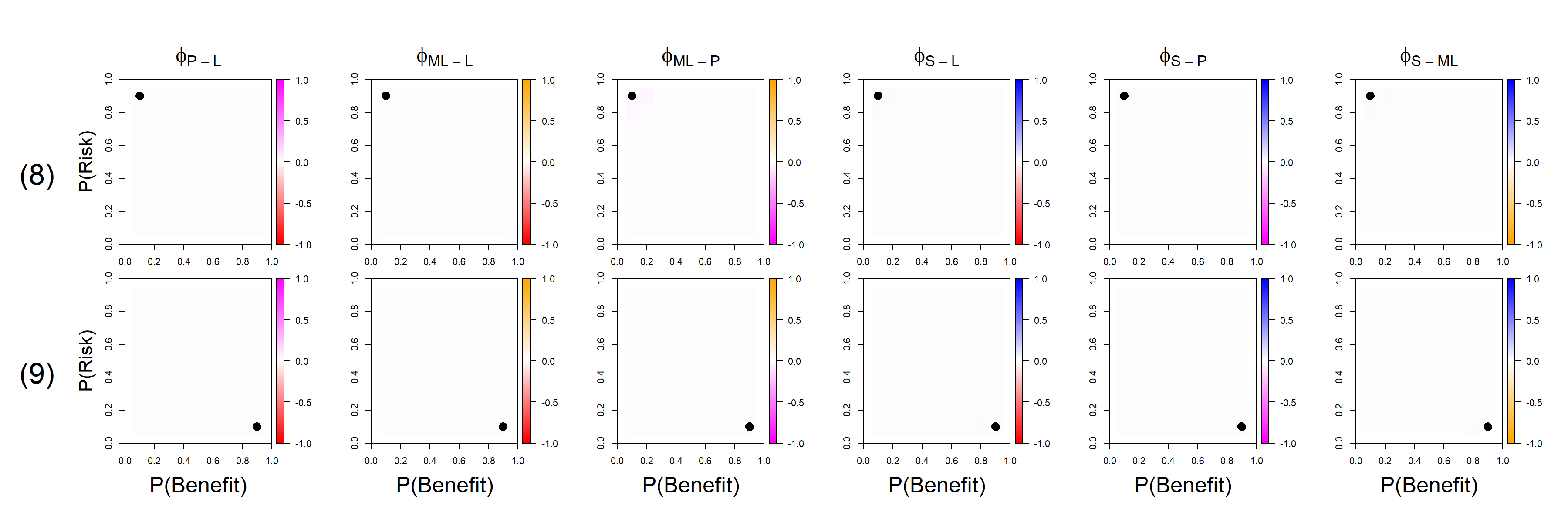}
\end{center}
\vspace{-2em}
\caption{Results of the six pairwise comparisons of the four AM, for scenarios 8 and 9}
\label{fig:sim3}
\end{figure}

\begin{table}[h]
\begin{center}
\noindent\makebox[\textwidth]{
\begin{tabular}{|cc|cccc|}
\hline
    & &Linear Model & Product Model & Multi-Linear Model & Product Model \\
\hline
Scenario 1    &     \begin{tabular}{@{}c@{}}$\geq$2.5$\%$ \\ $\geq$5$\%$\end{tabular} 
                & \begin{tabular}{@{}c@{}}24/81 \\ 12/81\end{tabular} 
                & \begin{tabular}{@{}c@{}}18/81 \\ 8/81\end{tabular} 
                & \begin{tabular}{@{}c@{}}25/81 \\ 9/81\end{tabular} 
                & \begin{tabular}{@{}c@{}}18/81 \\ 5/81\end{tabular} \\
\hline                
Scenario 2    &     \begin{tabular}{@{}c@{}}$\geq$2.5$\%$ \\ $\geq$5$\%$\end{tabular} 
                & \begin{tabular}{@{}c@{}}13/81 \\ 3/81\end{tabular} 
                & \begin{tabular}{@{}c@{}}18/81 \\ 1/81\end{tabular} 
                & \begin{tabular}{@{}c@{}}16/81 \\ 3/81\end{tabular} 
                & \begin{tabular}{@{}c@{}}16/81 \\ 1/81\end{tabular} \\
\hline
Scenario 3    &     \begin{tabular}{@{}c@{}}$\geq$2.5$\%$ \\ $\geq$5$\%$\end{tabular} 
                & \begin{tabular}{@{}c@{}}19/81 \\ 2/81\end{tabular} 
                & \begin{tabular}{@{}c@{}}13/81 \\ 2/81\end{tabular} 
                & \begin{tabular}{@{}c@{}}19/81 \\ 2/81\end{tabular} 
                & \begin{tabular}{@{}c@{}}13/81 \\ 0/81\end{tabular} \\
\hline
Scenario 4    &     \begin{tabular}{@{}c@{}}$\geq$2.5$\%$ \\ $\geq$5$\%$\end{tabular} 
                & \begin{tabular}{@{}c@{}}11/81 \\ 2/81\end{tabular} 
                & \begin{tabular}{@{}c@{}}0/81  \\ 0/81\end{tabular} 
                & \begin{tabular}{@{}c@{}}9/81  \\ 2/81\end{tabular} 
                & \begin{tabular}{@{}c@{}}0/81  \\ 0/81\end{tabular} \\
\hline
Scenario 5    &     \begin{tabular}{@{}c@{}}$\geq$2.5$\%$ \\ $\geq$5$\%$\end{tabular} 
                & \begin{tabular}{@{}c@{}}13/81 \\ 2/81\end{tabular} 
                & \begin{tabular}{@{}c@{}}3/81  \\ 0/81\end{tabular} 
                & \begin{tabular}{@{}c@{}}14/81 \\ 0/81\end{tabular} 
                & \begin{tabular}{@{}c@{}}0/81  \\ 0/81\end{tabular} \\
\hline
Scenario 6    &     \begin{tabular}{@{}c@{}}$\geq$2.5$\%$ \\ $\geq$5$\%$\end{tabular} 
                & \begin{tabular}{@{}c@{}}20/81 \\ 3/81\end{tabular} 
                & \begin{tabular}{@{}c@{}}13/81 \\ 0/81\end{tabular} 
                & \begin{tabular}{@{}c@{}}17/81 \\ 3/81\end{tabular} 
                & \begin{tabular}{@{}c@{}}8/81  \\ 0/81\end{tabular} \\
\hline
Scenario 7    &     \begin{tabular}{@{}c@{}}$\geq$2.5$\%$ \\ $\geq$5$\%$\end{tabular} 
                & \begin{tabular}{@{}c@{}}22/81 \\ 3/81\end{tabular} 
                & \begin{tabular}{@{}c@{}}14/81 \\ 2/81\end{tabular} 
                & \begin{tabular}{@{}c@{}}23/81 \\ 3/81\end{tabular} 
                & \begin{tabular}{@{}c@{}}11/81 \\ 0/81\end{tabular} \\
\hline
Scenario 8    &     \begin{tabular}{@{}c@{}}$\geq$2.5$\%$ \\ $\geq$5$\%$\end{tabular} 
                & \begin{tabular}{@{}c@{}}1/81 \\ 1/81\end{tabular} 
                & \begin{tabular}{@{}c@{}}2/81 \\ 1/81\end{tabular} 
                & \begin{tabular}{@{}c@{}}1/81 \\ 1/81\end{tabular} 
                & \begin{tabular}{@{}c@{}}3/81 \\ 1/81\end{tabular} \\
\hline
Scenario 9    &     \begin{tabular}{@{}c@{}}$\geq$2.5$\%$ \\ $\geq$5$\%$\end{tabular} 
                & \begin{tabular}{@{}c@{}}3/81 \\ 1/81\end{tabular} 
                & \begin{tabular}{@{}c@{}}3/81 \\ 1/81\end{tabular} 
                & \begin{tabular}{@{}c@{}}3/81 \\ 1/81\end{tabular} 
                & \begin{tabular}{@{}c@{}}3/81 \\ 1/81\end{tabular} \\

\hline
Total&   \begin{tabular}{@{}c@{}}$\geq$2.5$\%$ \\ $\geq$5$\%$\end{tabular} 
                & \begin{tabular}{@{}c@{}}126/729 \\ 29/729\end{tabular} 
                & \begin{tabular}{@{}c@{}}84/729  \\ 15/729\end{tabular} 
                & \begin{tabular}{@{}c@{}}127/729 \\ 24/729\end{tabular} 
                & \begin{tabular}{@{}c@{}}72/729  \\ 8/729\end{tabular}  \\
\hline
\end{tabular}
}\caption{Number of times when the difference in recommending $T_{1}$ changes by at least 2.5$\%$ or 5$\%$  between the negatively correlated criteria and the non-correlated criteria}
\vspace{-2em}
\label{fig:ncorr}
\end{center}
\end{table}

\newpage
\section{Weight derivations}

\subsubsection*{Product Model}
Firstly, note that under the product model, the resultant utility score for the two criteria example is given by:
\begin{equation*}
z^{P}= u^{P}(\theta_{i1},\theta_{i2}.w^{P})  =  u_{1}(\theta_{i1})^{w^{p}} \times u_{2}(\theta_{i2})^{1-w^{p}}
\label{eq:LP1}
\end{equation*}
This is then used to find the equation of the contour lines:
\begin{equation*}
1-u_{2}(\theta_{i2})= 1- \Bigg( \frac{z^{P}}{u_{1}(\theta_{i1})^{w^{p}}} \Bigg)^{\frac{1}{1-w^{P}}} \text{ ~for }u_{1}(\theta_{i1})^{w^{p}} > (z^{P})^{-w^{P}}
\label{eq:LP2}
\end{equation*}
The slope of the tangent of the contour at a given $u_{1}(\theta_{i1})$ for the utility score $z^{P}$  is:
\begin{equation*}
\frac{w^{P}}{1-w^{P}} \Bigg( \frac{1}{u_{1}(\theta_{i1})} \Bigg) \Bigg( \frac{z^{P}}{u_{1}(\theta_{i1})^{w^{p}}} \Bigg)^{\frac{1}{1-w^{P}}} 
\label{eq:LP3}
\end{equation*}
Now, the two equations of the slopes for the product and the linear models are set equal to each other at the point $u_{i,1}$=$u_{i,2}$=0.5. This gives the equality:
\begin{equation*}
\frac{w^{P}}{1-w^{P}}= \frac{w^{L}}{1-w^{L}}.
\label{LP4}
\end{equation*}

Which further simplifies down to $w^{P}$ =$w^{L}$ , showing that the weight used in the linear model is also the weight used in the product model.

\subsubsection*{Multi-Linear Model}

Firstly, note that under the multi-linear model, the resultant utility score for the two criteria example is given by:
\begin{equation*}
z^{ML}(\theta_{i1},\theta_{i2},w_{1}^{{ML}},w_{2}^{{ML}}):= w_{1}^{{ML}}u_1(\theta_{i1}) + (w_{2}^{{ML}})u_2(\theta_{i2}) + (1-w_{1}^{{ML}}-w_{2}^{{ML}})( u_1(\theta_{i1})u_2(\theta_{i2})).
\label{eq:LM1}
\end{equation*}
This is then used to find the equation of the contour lines:
\begin{equation*}
1-u_{2}(\theta_{i2})= 1- \frac{z^{ML}-w_{1}^{{ML}}u_1(\theta_{i1})}{w_{2}^{{ML}} + u_1(\theta_{i1}) - w_{1}^{{ML}}u_1(\theta_{i1}) - w_{2}^{{ML}}u_1(\theta_{i1}) }
\label{eq:LM2_1}
\end{equation*}
The slope of the tangent of the contour at a given $u_{1}(\theta_{i1})$ for the utility score $z^{S}$  is:
\begin{equation*}
\begin{array}{ccc}
\frac{w^{{L}}}{1-w^{{L}}} & = & \frac{w_{1}^{ML}w_{2}^{ML}+z^{ML}-z^{ML}(w_{1}^{ML}+w_{2}^{ML})}{\big{(}w_{2}^{ML}-(u_{1}(\theta_{i,1}))(-1+w_{1}^{ML}+w_{2}^{ML})\big{)}^2}
\end{array}
\label{eq:LM3_1}
\end{equation*}
Now, we firstly set a constraint $1-w_{1}^{ML}-w_{2}^{ML}=c$ (where $c$ is the total weight given to the interaction terms). We then set the two equations of the slopes for the multi-linear and the linear models so they are equal to each other at the point $u_{i,1}$=$u_{i,2}$=0.5. This gives the equality:
\begin{equation*}
{w^{ML}} = w^{L} - \frac{c}{2}.
\label{LM4_1}
\end{equation*}

\subsubsection*{SLoS Model}
Firstly, note that under the SLoS Model, the resultant loss score for the two criteria example is given by:
\begin{equation*}
z^{S} =\Bigg( \frac{1}{u_{1}(\theta_{i1})} \Bigg)^{w^{S}}  + \Bigg( \frac{1}{u_{2}(\theta_{i2})} \Bigg)^{1-w^{S}} 
\label{eq:LS1}
\end{equation*}
This is then used to find the equation of the contour lines:
\begin{equation*}
1-u_{2}(\theta_{i2})= 1- (z^{S}-u_{1}(\theta_{i1})^{-w^{S}})^{-\frac{1}{1-w^{S}}} \text{ ~for }u_{1}(\theta_{i1}) > z^{-\frac{1}{z^{S}}}
\label{eq:LM2_2}
\end{equation*}
The slope of the tangent of the contour at a given $u_{1}(\theta_{i1})$ for the utility score $z^{S}$  is:
\begin{equation*}
\frac{w^{S}}{1-w^{S}} \bigg( z^{S}-u_{1}(\theta_{i1})^{-w^{S}} \bigg)^{\frac{w^{s}-2}{1-w^{S}}} \bigg( u_{1}(\theta_{i1})^{-(w^{S}+1)} \bigg) 
\label{eq:LM3_2}
\end{equation*}
Now, the two equations of the slopes for the SLoS and the linear models are set equal to each other at the point $u_{i,1}$=$u_{i,2}$=0.5. This gives the equality:
\begin{equation*}
\frac{w^{S}}{1-w^{S}} \bigg(2^{2w^{s}-1} \bigg)= \frac{w^{L}}{1-w^{L}}.
\label{LM4_2}
\end{equation*}